\documentclass[reprint,amsmath,amssymb,aps,pre,a4paper,twoside]{revtex4-2}
\usepackage{graphicx}
\usepackage{dcolumn}
\usepackage{bm}
\usepackage[utf8]{inputenc}
\usepackage[T1]{fontenc}
\usepackage{mathptmx}
\usepackage{subcaption}
\DeclareGraphicsExtensions{.png,.jpg,.jpeg}
\usepackage{hyperref}
\usepackage{caption}
\usepackage{amsmath}
\usepackage{soul}
\usepackage{color}
\usepackage{float}
\usepackage{placeins}

\begin{abstract}
	Studies of transient dynamics captures the time history of any dramatic changes in the dynamics of a system. The transient dynamics is investigated here in a classic ecological model of a bistable tri-trophic food chain. All the species in the food chain either coexist or undergoes a partial extinction with the loss of predator after a transient time that depends upon the initial density of species populations. The distribution of transient time of extinction of the predator in response to initial states shows interesting pattern of inhomogeneity and anisotropy in the basin of predator-fee state. Precisely, the distribution shows a multimodal character when the initial points are near the basin boundary, and unimodal at locations far from the border. The distribution is also anisotropic in the sense that the number of modes depends on the direction from the basin location. We define two new metrics, viz., homogeneity index and local isotropic index to characterize the distinctive features of the distribution. The inhomogeneity of distribution reduces significantly with resource enrichment, however, a similar change in anisotropy is relatively low. We try to explain the origin of such multimodal distributions from the dynamical system perspectives and present its ecological implications.
\end{abstract}

\begin{document}
	
\title{Multimodal distribution of transient time of predator extinction in a three-species food chain }
\preprint{APS/123-QED}

\author{Debarghya Pattanayak$^{1}$, Arindam Mishra$^{2}$, Syamal K. Dana$^{1,2,3}$, Nandadulal Bairagi$^{1}$}
\address{$^{1}$ Centre for Mathematical Biology and Ecology, Department of Mathematics, Jadavpur University, Kolkata 700032, India.\\ $^{2}$ Division of Dynamics, Lodz University of Technology, Stefanowskiego 1/15, 90-924 Lodz, Poland \\$^{3}$ Department of Mathematics, National Institute of Technology, Durgapur 713209, India.}

\thanks{Corresponding author: nbairagi.math@jadavpuruniversity.in}

\date{\today}
\maketitle
\section{Introduction}
\par $\;$ $\;$ A study of asymptotic behavior gives a partial picture of dynamics of a system \cite{hastings2018transient, lai2011transient}. The behavior of a system for the intermediate time remains unknown before it reaches the asymptotic state. In particular, in ecological systems, a study of only the asymptotic behavior, therefore, may hinder any attempt to exercise any form of control on the system's behavior at its initial stage, that shall beneficially affect its long-term behavior, such as inducing  any support for survival to the vulnerable species in the transient state, which may lead to the coexistence of all the species in the ecosystem \cite{dhamala1999controlling}. Moreover, ignoring the underlying dynamics that a system may exhibit from its initial point to its long-term behavior keeps us oblivious to the nature of the evolution of the system. The dynamics that exists before reaching an asymptotic state is called the transient dynamics, and the time required for any trajectory of a system to reach the final attractor (asymptotic state) from the initial state is called the transient time \cite{grebogi1986critical,yorke1979metastable, ray2021mitigating}. In recent years, the statistics analysis of transient time has received a considerable attention due to its relevance in complex systems \cite{altmann2013leaking, lilienkamp2017features,lilienkamp2018terminal}, climatic models \cite{lenton2011early, scheffer2009early}, ecology \cite{morozov2020long,gosztolai2019collective,martin2020importance, arnoldi2018ecosystems, gao2016universal} and dynamical networks \cite{hens2019spatiotemporal, tarnowski2020universal}, in general. The statistics of transient time is important for  predicting  extreme events \cite{majumdar2020extreme}, species extinction \cite{dhamala1999controlling, hastings2004transients, hastings2010timescales, hastings2018transient} and controlling power grid failure \cite{motter2013spontaneous}. The role of transient time becomes all the more profound for assessing the critical transitions to alternate stable states or predicting a tipping point \cite{vanselow2019very}.

\par The statistics of transient time is more complicated for multistable systems \cite{feudel2008complex}, where the choice of initial states affects the transient dynamics. Hence, the statistics of transient time needs to be assessed against the pertaining sub-basins \cite{nusse1996basins} of the multistable system. The challenges posed to such analysis become all the more aggravated when multimodal distribution of transient time arises, making our conventional statistical approaches of mean, median, and other central tendencies difficult to apply. In the event of multiple modes in the distribution, neither following any known patterns, normal, Poisson, nor the modes being equispaced, the only choice left with us is the  analysis of each distinct mode in the multimodal distribution. Moreover, when dealing with such multistable systems displaying multimodal distributions, the observation of the variation in the number of distinct modes, along with other statistical parameters, simulated with initial points at different distances and on different directions of the basin boundary, become a matter of utmost concern. If any noticeable trends are found in the obtained results, attempts could be made to characterize the structure of the basin from the perspective of transient time distributions.

\par We address the problem of analyzing the multimodal distribution of transient time that appears in a paradigmatic tri-trophic food chain model \cite{yodzis1992body, hastings1991chaos,mccann1994nonlinear}, which is well known for its bistability. The system either shows coexistence of all the species or a predator-free limit cycle depending on the initial population size and parameter values. What has not been explored so far, in this bistable system, is the distribution of transient time of the predator's extinction. The focal point of our study is to examine the distribution of this transient time that appears to be multimodal, and to provide some techniques for its analysis. We adopt a method of partitioning the concerned basin of attraction into identical blocks of reasonably small size and then analyze the transient time and its distributions in each block, separately, and subsequently build a larger picture of the entire distribution scenario of the system. During this process, efforts are also given to characterize the basin structure of the model system from the perspective of distribution of the transient time, which appears to be inhomogeneous and anisotropic. Broadly speaking, the number of modes varies with the distance of the initial states from the basin boundary and also depends upon the direction of the initial points relative to the same boundary.

\par The organization of the paper is as follows. A brief introduction of the model and its bistability with the basin of attraction are presented in Section \ref{food chain model}. Section \ref{distribution of transient time} discusses the trends in the distribution of transient time  of extinction of the predator and the reason behind the existence of such multimodal distributions. In Section \ref{basin sturcture}, a description is given how the basin of predator-free limit cycle is partitioned into small identical blocks and the multimodal distribution is derived. Two new metrics are defined to quantify the structural pattern of the basin with respect to the number of modes in the transient time distributions. Subsequently, attributes of the distributions in terms of the new metrics are discussed to explain the noted multimodal distribution pattern. The paper ends with a brief discussion in Section \ref{discussion}.

\section{Tri-tropic food chain model}\label{food chain model}
\par $\;$  $\;$ The tri-trophic food chain \cite{rosenzweig1973exploitation, hastings1991chaos, yodzis1992body} presents the interactions between a resource, a consumer and a predator. The model assumes that the resource grows in a logistic fashion in the absence of any consumer, whereas the consumer and predator both exponentially decay in the absence of their respective prey. The consumer and predator have saturated Holling Type-II functional responses, and there are no losses of time and energy during hunting and digestion. Three species grow simultaneously and have overlapping generations with no age structure, where the growth rates of resource population density ($R$), consumer population density ($C$) and predator population density ($P$) at time $t$ are defined by 
\begin{equation}\label{Model}
\begin{split}
\frac{dR}{dt} & = R(1-\frac{R}{K})-x_{C}y_{C}\frac{CR}{R+R_0},\\
\frac{dC}{dt} & = -x_{C}C(1-y_{C}\frac{R}{R+R_0})-x_{P}y_{P}\frac{PC}{C+C_0},\\ 
\frac{dP}{dt} & = -x_{P}P(1-y_{P}\frac{C}{C+C_0}).
\end{split}
\end{equation} 
,where $R(0),C(0),P(0)>0$.\\
The default parameter values as used for all the simulations here, in this study, are taken from McCann and Yodzis \cite{mccann1994biological} and shown in Table \ref{T1}. The parameter values are biologically plausible with a feasible prey-predator body mass ratio and consistent with the consumer or predator being a vertebrate or invertebrate ectotherm \cite{mccann1994biological}.
\begin{table*}
	\centering
	\captionof{table}{Parameters description and their default values \cite{mccann1994biological}.} \label{T1}
	\begin{tabular}{| m{2cm} | m{25em}| m{1cm} | } 
		\hline 
		Parameter & Description & Value\\
		\hline \hline
		$K$ & Resource carrying capacity & 0.95 \\ 
		\hline
		$R_{0}$ & Resource half-saturation density & 0.161 \\
		\hline
		$C_{0}$ & Consumer half-saturation density & 0.5 \\
		\hline
		$x_C$ & Energy consumption rate of $C$ per unit body weight relative to the rapidity with which the resource replaces itself & 0.4\\
		\hline
		$x_P$ & Energy consumption rate of $P$ per unit body weight relative to the rapidity with which the resource replaces itself & 0.08\\
		\hline
		$y_C$ & Energy spent per unit energy acquired by $C$ & 2.01\\
		\hline
		$y_P$ & Energy spent per unit energy acquired by $P$ & 2.876\\
		\hline		    
	\end{tabular}
\end{table*}
\par
\begin{figure}[h!]
	\hspace*{-4.5cm}
	\begin{subfigure}[h!]{0.23\textwidth}
		\centering
		\subcaptionbox{ }
		{\includegraphics[width=2\textwidth,height=5.25cm]{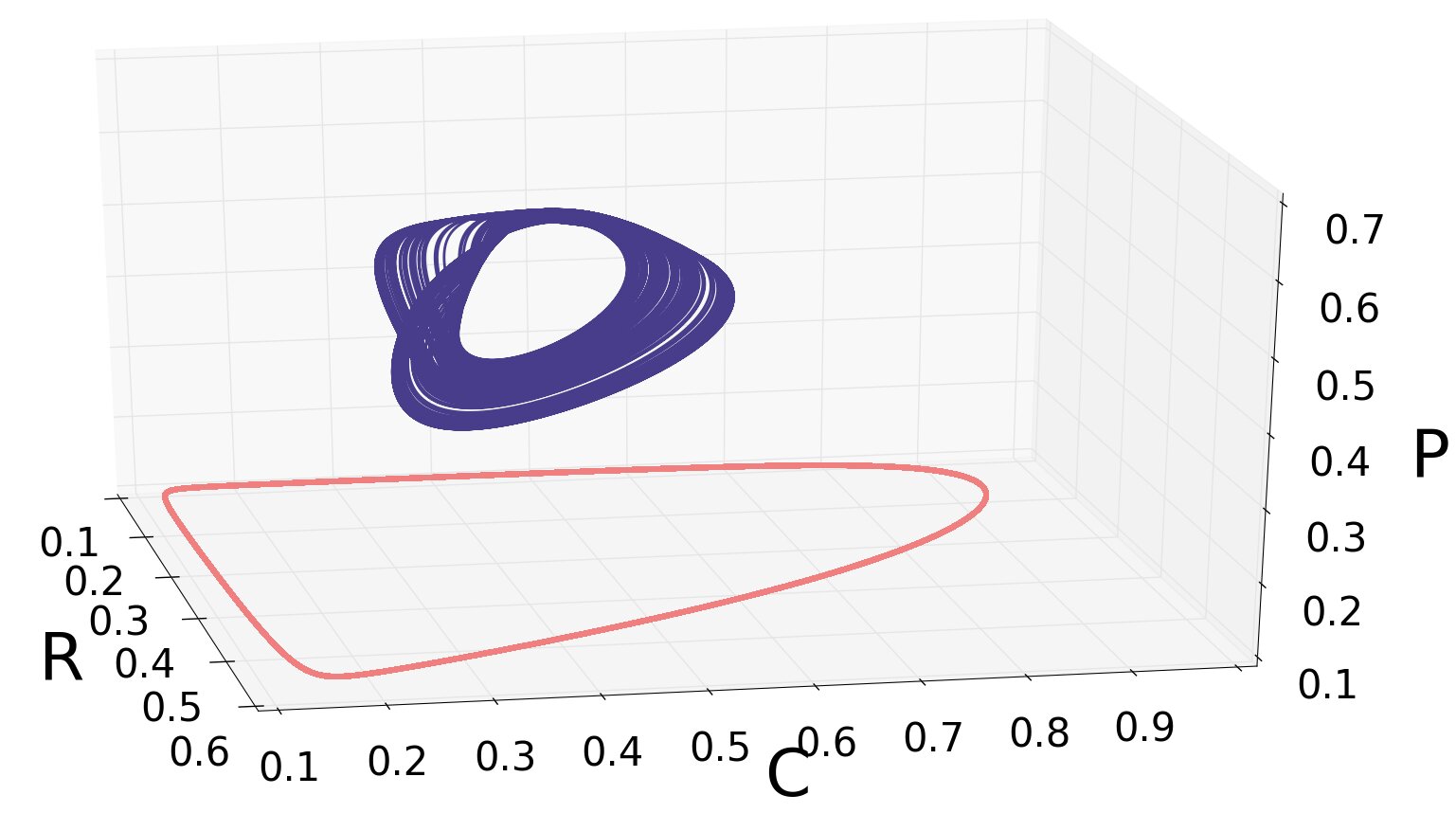}%
		}\qquad		
	\end{subfigure}\\
	\begin{subfigure}[h!]{0.23\textwidth}
		\hspace*{-3cm}
		\centering
		\subcaptionbox{ }
		{\includegraphics[width=2.25\textwidth,height=5.25cm]{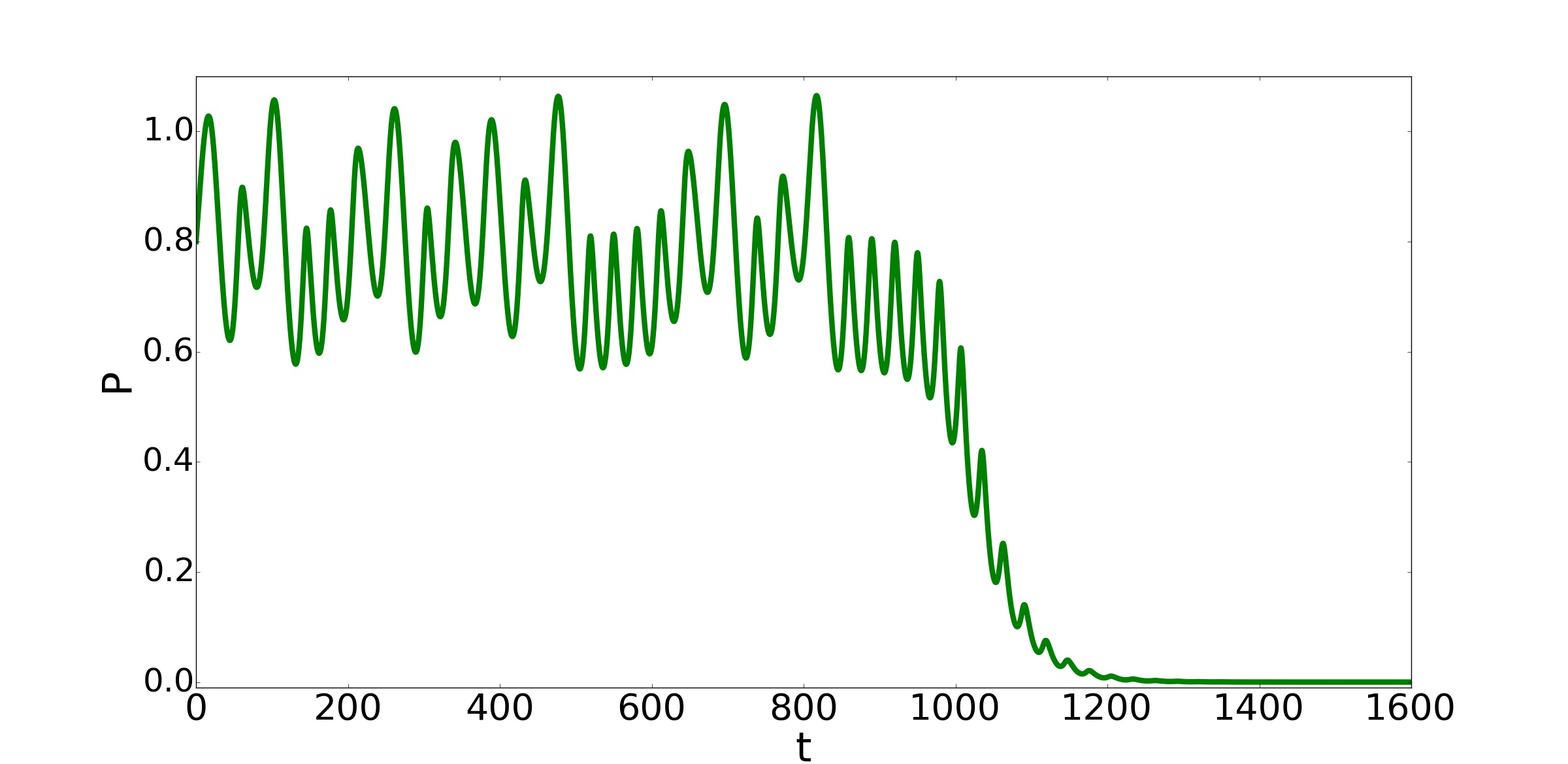} %
		}\qquad	
	\end{subfigure}
	\caption{{\bf Bistability and transient dynamics of the 3-species food chain.} (a) A chaotic attractor (blue) coexists with a predator-free period-1 limit cycle (red line) in $R-C-P$ state space. (b) Transient dynamics of the predator population is shown with its extinction when the system transits to an asymptotic state of limit cycle oscillation ($P=0$). }
	\label{Bistability}		
\end{figure}

 $\;$ This classic food chain model has been elaborately investigated as reported earlier  \cite{pattanayak2021bistability, mccann1994nonlinear} and it is bistable for a broad range of parameters where two dynamical regimes exist with their distinct basins separated by a smooth boundary. If the initial population densities of species are taken from one of the basins, they lead to a coexisting attractor \cite{mccann1994nonlinear} depending upon the choice of parameters. For an alternative choice of initial states from the other basin, the trajectories will enter the predator-free period-1 limit cycle. A prototype illustration of bistability and transient dynamics of predator extinction as typically seen in the model is presented in Fig.~\ref{Bistability} for a selective choice of parameters. 

\subsection{Distribution of transient time}\label{distribution of transient time}
\par
\begin{figure*}[ht!]
	\hspace*{-4cm}
	\begin{subfigure}[b]{0.2\textwidth}
		\centering
		\subcaptionbox{}{%
			\includegraphics[width=1.8\textwidth,height=5cm]{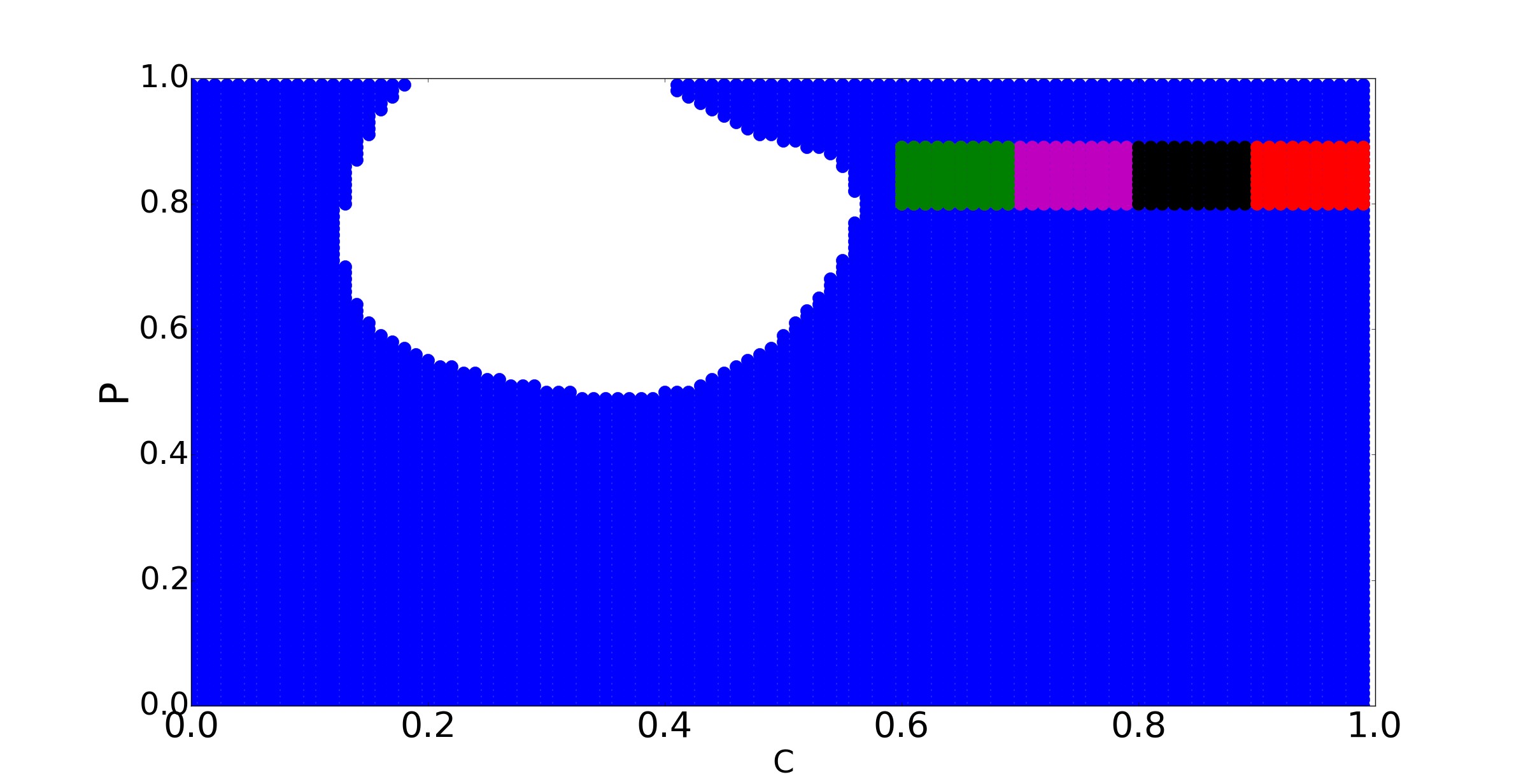} %
		}\qquad
	\end{subfigure}\hspace*{3cm}
	\begin{subfigure}[b]{0.2\textwidth}
		\centering
		\subcaptionbox{}{%
			\includegraphics[width=1.8\textwidth,height=5cm]{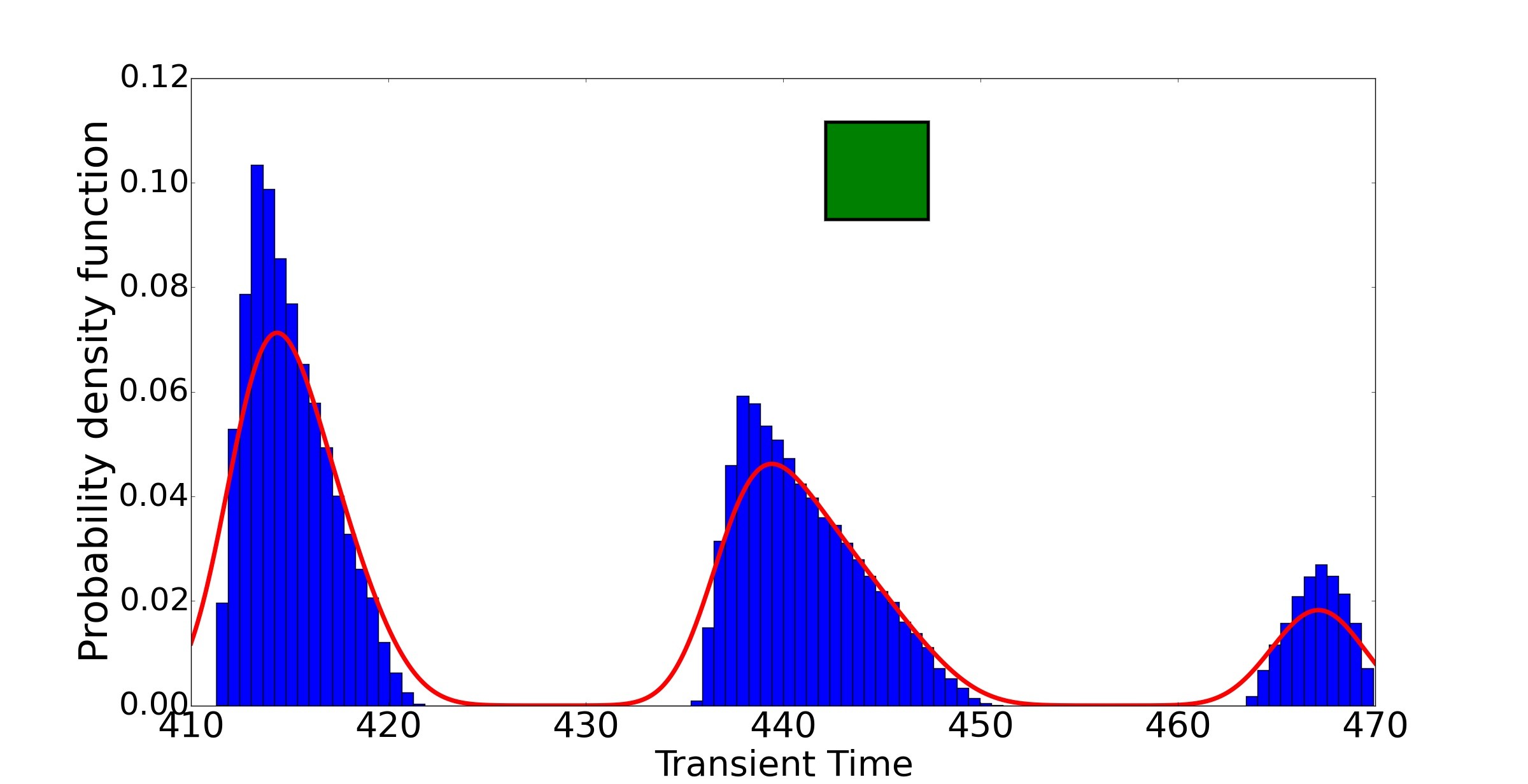}%
		}\qquad
	\end{subfigure}\\
	\hspace*{-4cm}
	\begin{subfigure}[b]{0.2\textwidth}
		\centering
		\subcaptionbox{}{%
			\includegraphics[width=1.8\textwidth,height=5cm]{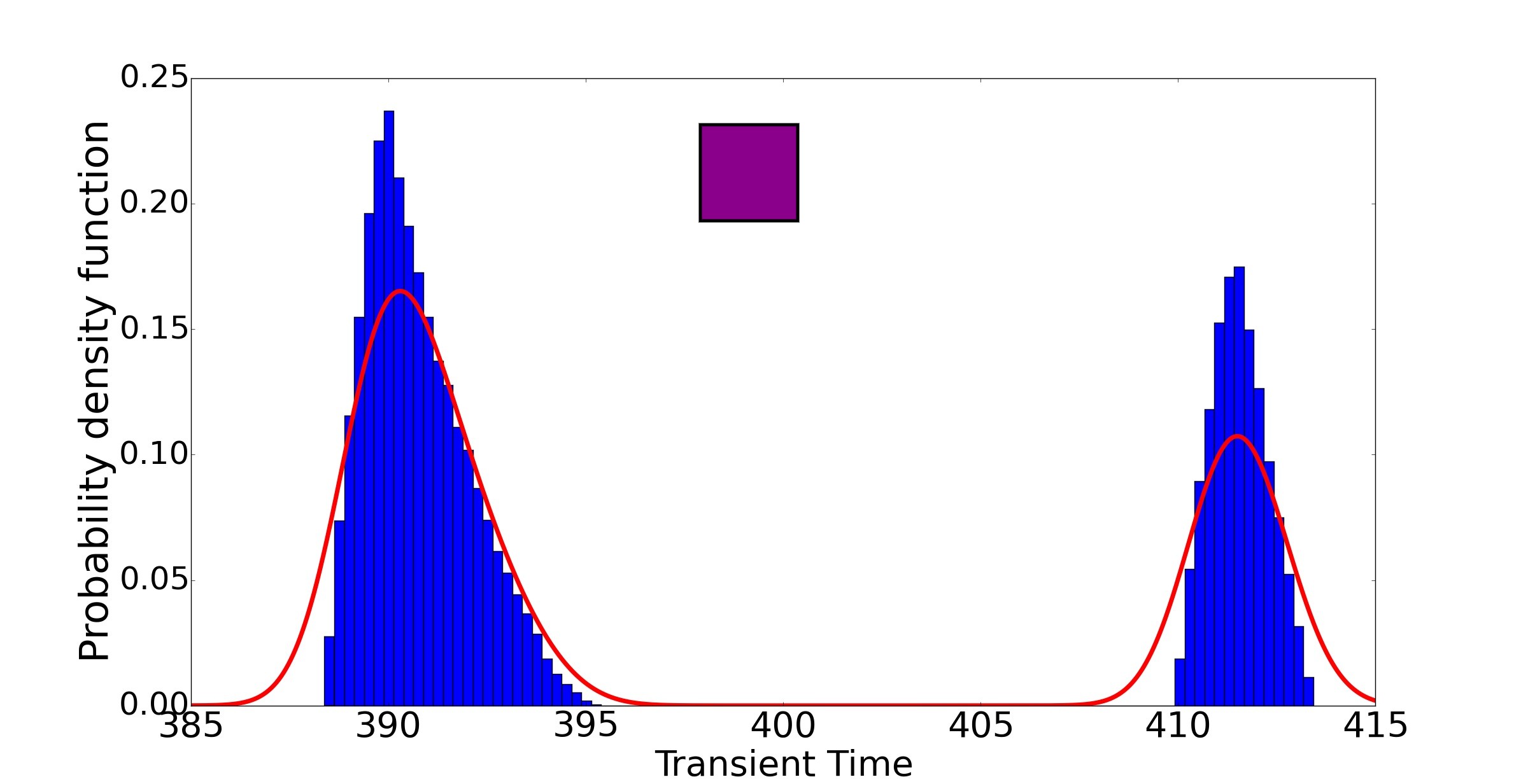}%
		}\qquad
	\end{subfigure}\hspace*{3cm}
	\begin{subfigure}[b]{0.2\textwidth}
		\centering
		\subcaptionbox{}{%
			\includegraphics[width=1.8\textwidth,height=5cm]{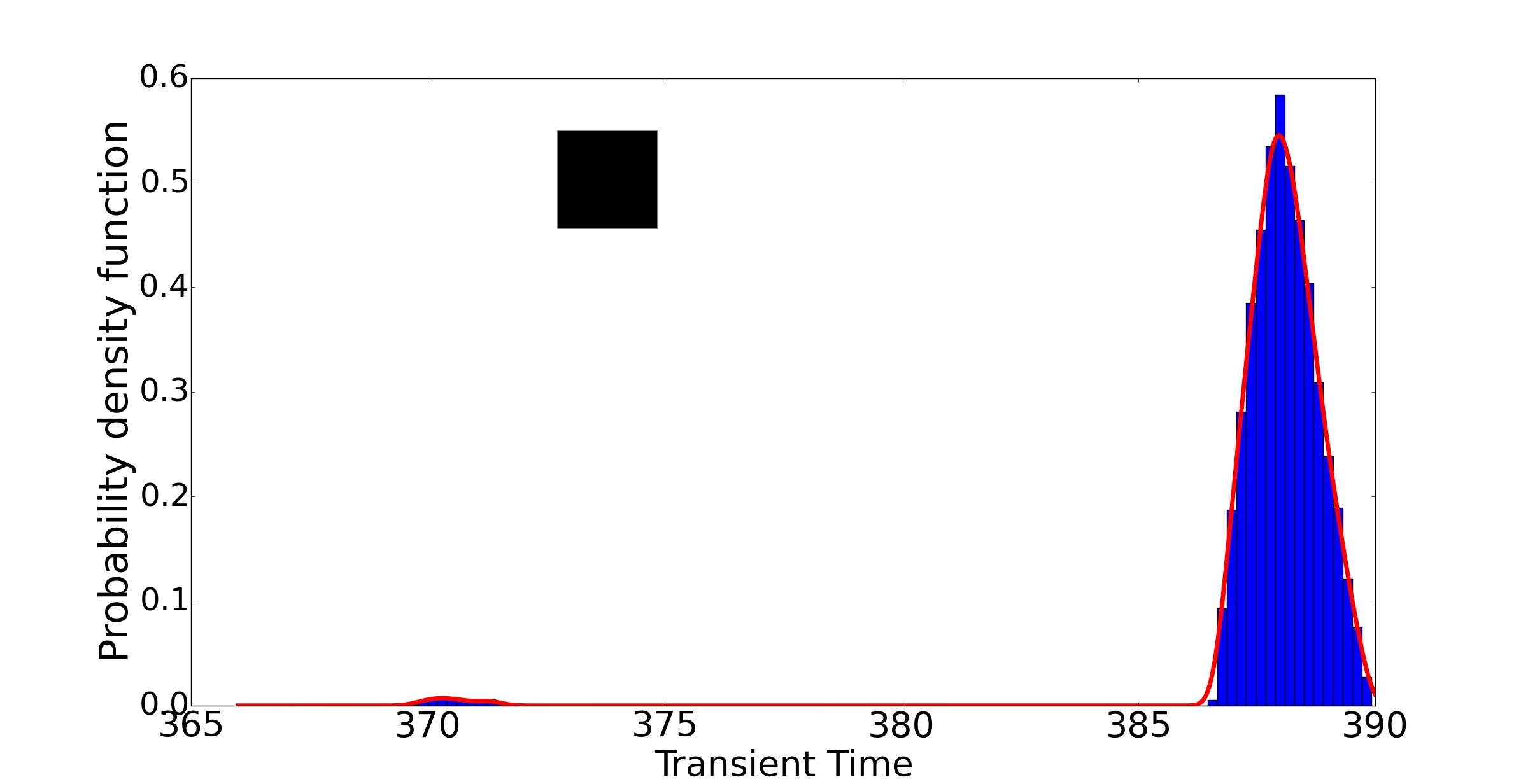}%
		}\qquad
	\end{subfigure}\\
	\hspace*{-4cm}
	\begin{subfigure}[b]{0.2\textwidth}
		\centering
		\subcaptionbox{}{%
			\includegraphics[width=1.8\textwidth,height=5cm]{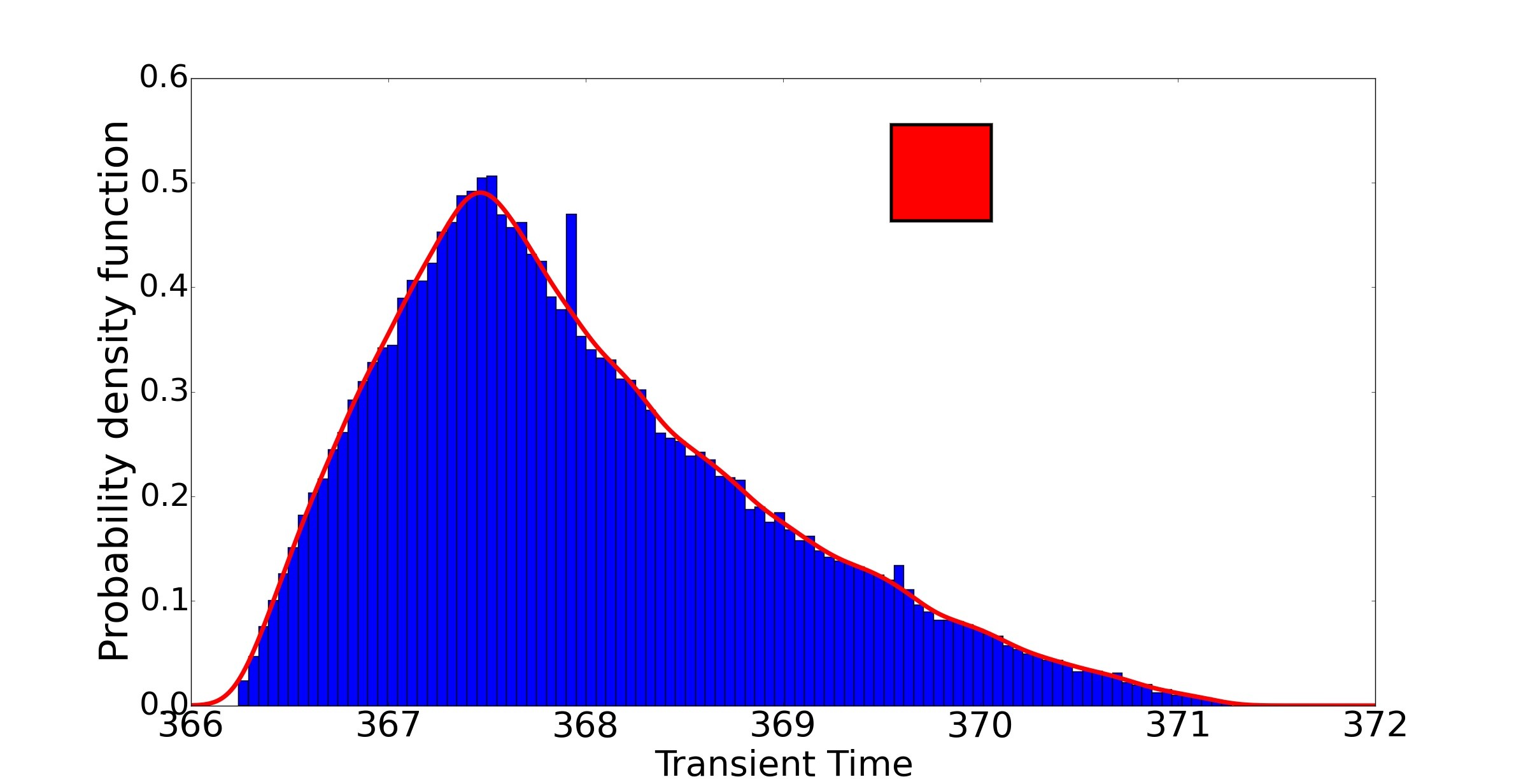}%
		}\qquad
	\end{subfigure}\hspace*{3cm}
	\begin{subfigure}[b]{0.2\textwidth}
		\centering
		\subcaptionbox{}{%
			\includegraphics[width=1.8\textwidth,height=5cm]{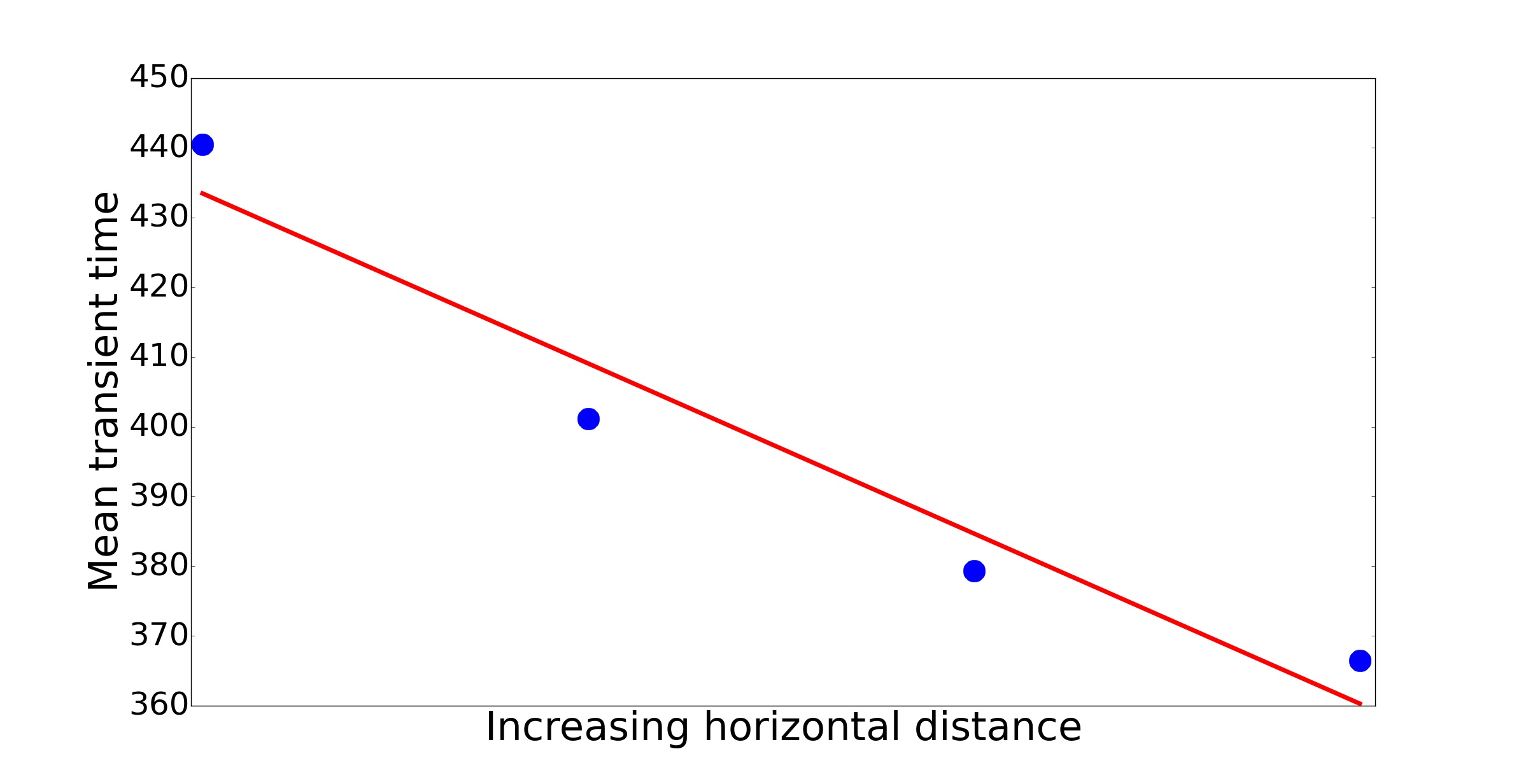}%
		}\qquad
	\end{subfigure}
	\caption{{\bf Basin of attraction of the bistable tri-tropic food chain and distribution of transient time.} (a) Basin of attraction in a $C-P$ plane with two distinct basins separated by a smooth boundary. White basin for coexisting species, and blue basin for predator-free limit cycle. Location of the green box ($[0.6,0.7]\times[0.8,0.9]$), magenta ($[0.7,0.8]\times[0.8,0.9]$), black ($[0.8,0.9]\times[0.8,0.9]$) and red ($[0.9,1.0]\times[0.8,0.9]$) boxes in the blue basin. Modes of distribution of transient time: (b) trimodal, (c) bimodal, (d) bimodal, and (e) unimodal. Red curves in (b, c, d, e) are the fitted lines of distributions, which are log-normal. (f) Mean transient time ($MTT$) calculated for multiple modes in each box, decreases with distance of the boxes from the basin boundary. The red line is the regression line. Parameter values are given in Table \ref{T1}.}
	\label{Distribution} 
\end{figure*}

 $\;$ $\;$ For a broad range of parameters, in the initial stages, the predator coexists in chaotic or periodic oscillations with resource and consumer and then experiences an extinction as shown in in Fig.~\ref{Bistability}(b). Fig.~\ref{Distribution}(a) shows the bistable basin, where the white basin corresponds to coexistence dynamics (periodic for this parameter set) of the three pertaining species and the blue basin represents the period-1 limit cycle when the predator goes to extinction after a transient time. We compute the distribution of transient time of predator extinction and search for its possible variation with the location of initial points. For this purpose, we make an preliminary check with a selection of four adjacent identical square blocks (green, magenta, black and red) of edge $10^{-1}$ units in the blue basin shown in Fig. \ref{Distribution}(a). The square blocks are arranged horizontally in a sequence of increasing distance from the basin boundary. Subsequently, a large number of initial points ($10^5$) are chosen randomly from each of the colored blocks and the transient time of the trajectory for all the selected initial points in each block is numerically estimated and their distribution is plotted. This is repeated for all the four colored boxes and plotted separately, when a distinct characteristic feature is obtained with a variation in the distribution of transient time in the blocks. A multimodal distribution of transients is observed and the number of modes varies for each block, along with a variation in the mean transient time. Figs.~\ref{Distribution}(b)-(e) present a picture of a decreasing number of modes in the distribution as we move from one block to another, away from the basin boundary to make our choices of initial points. This distinctive qualitative feature of distributions and the trend remains invariant with the number of initial points (checked with different number of points), provided they are chosen randomly and sufficiently large in numbers.

Figure \ref{Distribution}(b) is produced with initial points chosen from the block (green) at a grid location $[0.6,0.7]\times[0.8,0.9]$ closest to the basin boundary when three distinct modes are observed. As we move away from the basin boundary to the adjacent blocks (magenta and black) at grid locations $[0.7,0.8]\times[0.8,0.9]$ (magenta) and $[0.8,0.9]\times[0.8,0.9]$ (black), the distribution changes from tri-modal to bimodal as shown in Fig.~\ref{Distribution}(c) and Fig.~\ref{Distribution}(d), respectively. One mode for initial conditions from the black box is not so prominent, yet the distribution is clearly bi-modal, indicating the decreasing number of modes with the increasing distance from the basin boundary. The furthest block (red) at location $[0.9,1.0]\times[0.8,0.9]$ shows unimodal distribution of transient time of predator extinction in Fig.~\ref{Distribution}(e). The mean transient time ($MTT$) is plotted for each block in Fig.~\ref{Distribution}(f), and interestingly, it also decreases with the increasing distance of the boxes from the basin boundary. Focusing separately on each mode for all the blocks in Figs.~ \ref{Distribution}(b)-(e), we opine that each mode follows a log-normal distribution (fitted with red lines). The statistical properties such as mean, variance and deviations (fluctuations) in transient time for each block are presented in Table \ref{T2}.

\subsection{Origin of multimodal distribution}
\par
\begin{figure*}[ht!]
	\hspace*{-4cm}
	\begin{subfigure}[b]{0.24\textwidth}
		\centering
		\subcaptionbox{}{%
			\includegraphics[width=2\textwidth,height=5.5cm]{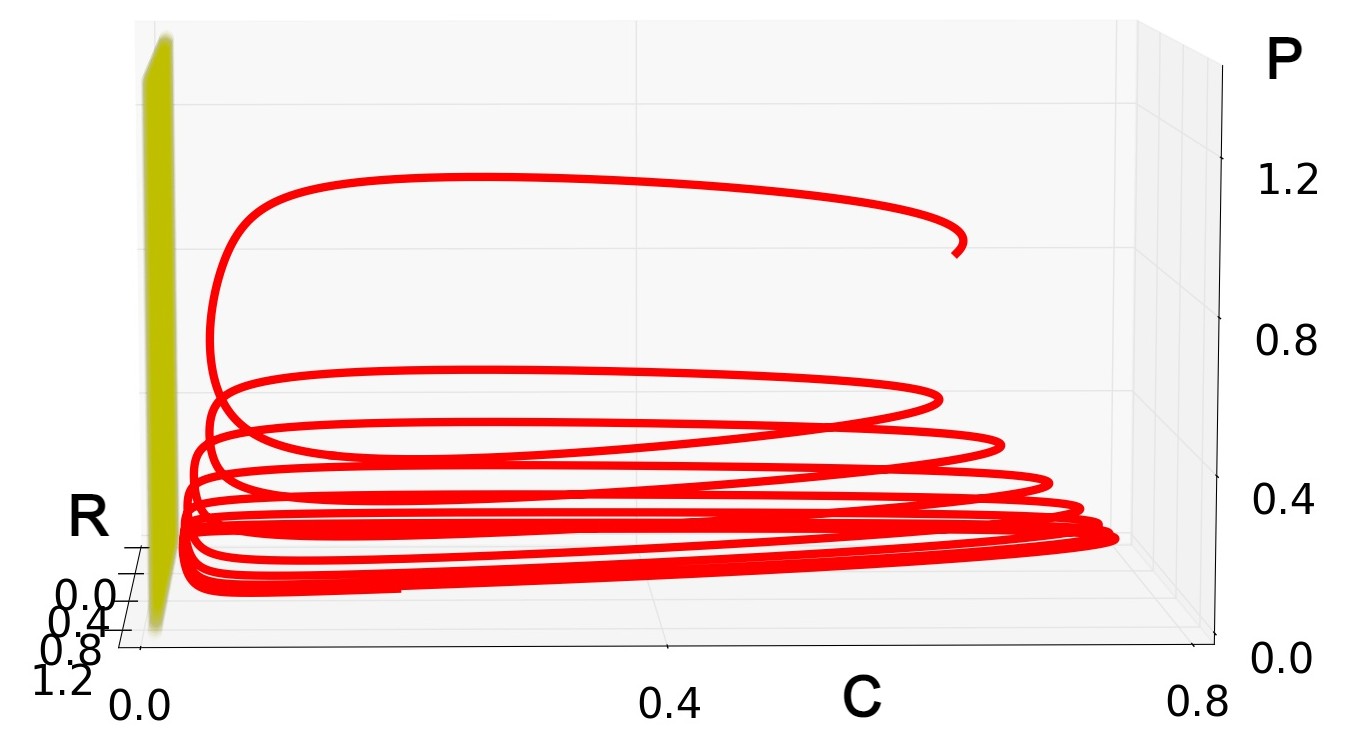} %
		}\qquad
	\end{subfigure}\hspace*{3.6cm}
	\begin{subfigure}[b]{0.24\textwidth}
		\centering
		\subcaptionbox{}{%
			\includegraphics[width=2\textwidth,height=5.5cm]{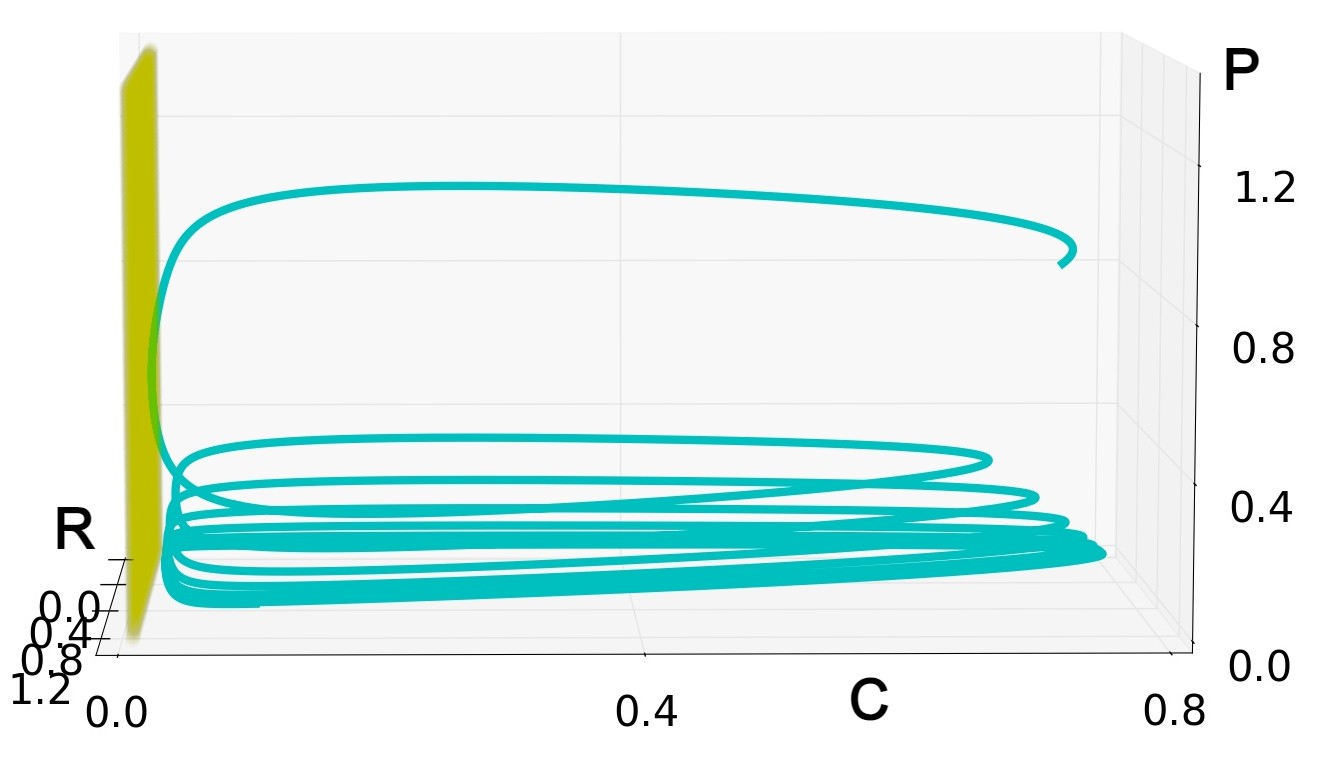}%
		}\qquad
	\end{subfigure}\\
	\hspace*{-4cm}
	\begin{subfigure}[b]{0.24\textwidth}
		\centering
		\subcaptionbox{}{%
			\includegraphics[width=2\textwidth,height=5.5cm]{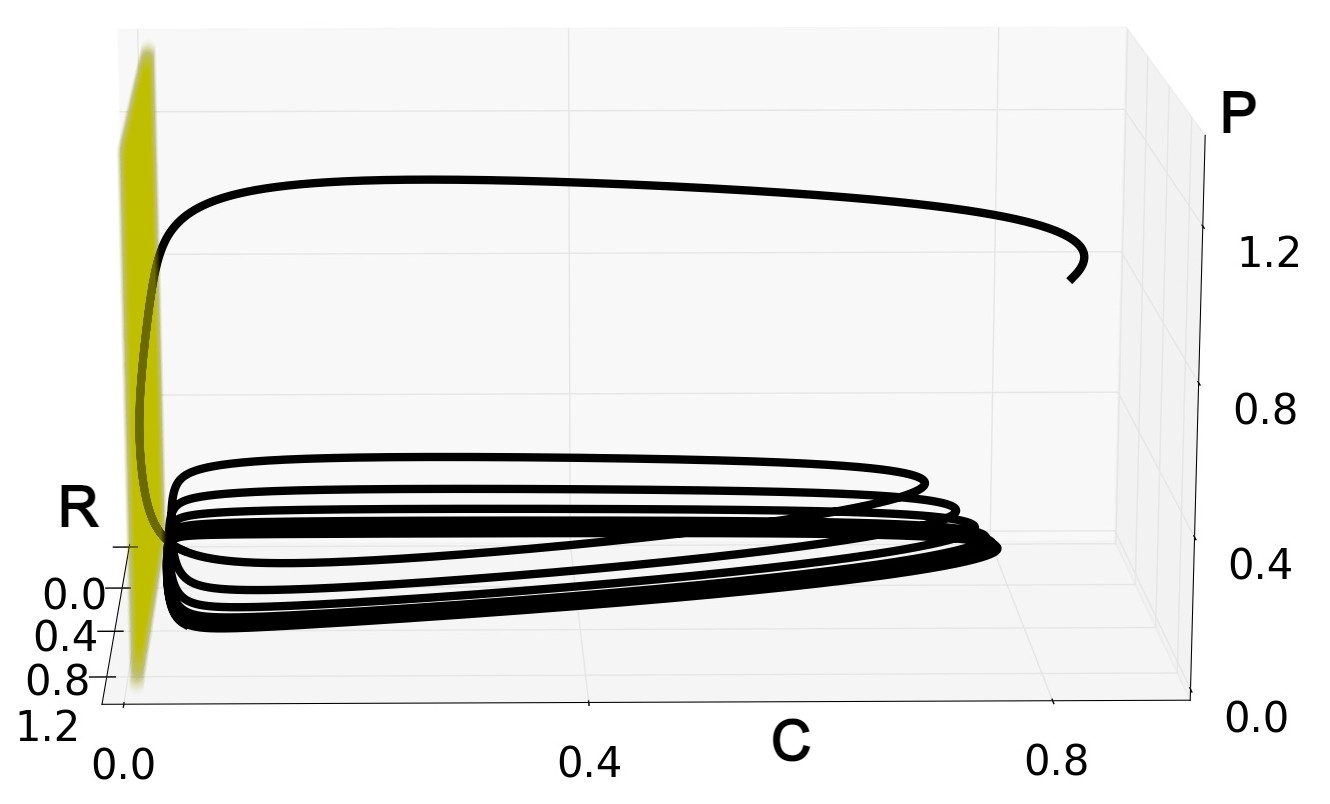}%
		}\qquad
	\end{subfigure}\hspace*{3.6cm}
	\begin{subfigure}[b]{0.24\textwidth}
		\centering
		\subcaptionbox{}{%
			\includegraphics[width=2\textwidth,height=5.5cm]{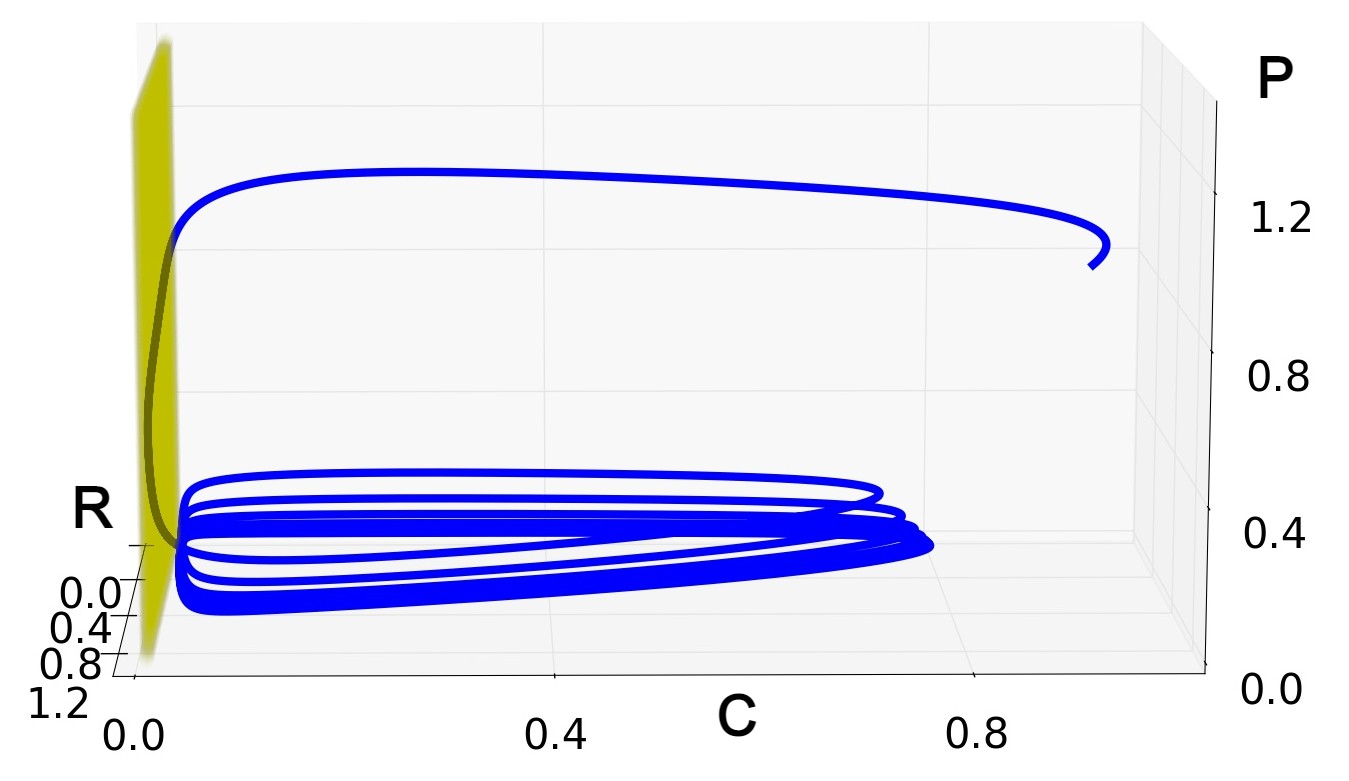}%
		}\qquad
	\end{subfigure}
	\caption{{\bf Trajectories of the tri-tropic food chain model.} 3-dimensional trajectories for a variety of choices of initial points from separate blocks showing different modes. Transient dynamics of the system when initial points are chosen from (a) green box at  (0.5,0.65,0.85) in Fig.~\ref{Distribution}(b), (b) magenta box at (0.5,0.75,0.85) in Fig. \ref{Distribution}(c), (c) black box at (0.5,0.85,0.85) in Fig. \ref{Distribution}(d) and (d) (0.5,0.95,0.85) from the red box in Fig. \ref{Distribution}(e), respectively. Exemplary trajectories for different choices of initial points show a decay of the predator population while grazing the light yellow plane at $C=0.01$, and then make a descent to the $P=0$ plane with a spiraling motion. Parameters are given in Table \ref{T1}.}
	\label{Mech} 
\end{figure*}

 $\;$ $\;$ 	We address an obvious question how the distinct modal distributions of transient time originate. Figures~\ref{Mech}(a)-(d) show exemplary trajectories of the 3-species food chain in $R-C-P$ space for different choices of initial states, taken from the middle of the four colored blocks in Fig.~\ref{Distribution}(a). The qualitative picture of a trajectory remains unchanged for varying choices of initial conditions within each box. It reveals that the system follows different pathways to arrive at a predator-free state, through spiraling for a longer run and alternatively, via grazing the plane $C=0.01$ (yellow-colored plane, as evident from our data) followed by a shorter run on spiraling. For illustrations, we consider four particular initial values, taken from each colored box, with only a change in the initial consumer density. For the initial state from the green box $(0.5, 0.65, 0.85)$, the trajectory spirally descends to reach the predator-free state ($P=0$) avoiding the yellow plane at $C=0.01$. For another set of initial points from the magenta box $(0.5, 0.75, 0.85)$, the trajectory comes closer to the plane $C=0.01$, in the beginning, yet not touching it as shown in Fig.~\ref{Mech}(b), before finally spiraling on to the $P=0$ plane of the stable limit cycle. The trajectories shown in Figs.~\ref{Mech}(c-d) are somewhat similar in nature with initial conditions taken from the black box $(0.5, 0.85, 0.85)$ and the red box $(0.5, 0.95, 0.85)$, respectively. In both the cases, the trajectory first grazes the $C=0.01$ plane for some duration of time and then make a short run of spiraling on before reaching the $P=0$ plane of the limit cycle.
\par  We denote the total transient time of any trajectory as $T$, counting from the initial state and the duration of reaching the predator-free limit cycle state. The entire transient time $T$ is divided into three time segments when the trajectory is assumed to spend (i) $T_1>0$ time to reach a close location of the ($C=0.01$, yellow) plane from the initial state, (ii) $T_2\geq 0$ time spent on ($C=0.01$, yellow) plane, (iii) $T_3>0$ time of spiraling before reaching the predator-free limit cycle. The total time $T$ taken by a trajectory to arrive at  the predator-free limit cycle from any initial states is
\begin{equation}\label{tt}
\begin{split}
T=T_1+T_2+T_3.
\end{split}
\end{equation}

\par  Figures~\ref{Mech} show that $T_1,T_3\neq0$, for all the choices of initial states, however, $T_2$=0 when the trajectory never grazes the $C=0.01$ plane. When the trajectories descent in such a pure spiral motion, avoiding the $C=0.01$ plane, the radius of evolution of a spiral gradually increases in each cycle before, finally, arriving at the asymptotic state of stable predator-free limit cycle. The average time required to complete one revolution while spiraling is assumed to be $T^{'}$. 

\par  For the first case in Fig. \ref{Mech}(a), we evaluate $T_1=24.79$, $T_2=0$ (since the trajectory avoids the $C=0.01$ plane never grazing it) and $T^{'}=27.7$. We use a Poincar\'e surface of section to estimate the time intervals $T_1$ and $T'$ (see Appendix). If $n\in Z_+$ denotes the total number of revolutions during spiraling before reaching the predator-free limit cycle, we obtain from Eq.~(\ref{tt})
\begin{equation}\label{tt1}
	\begin{split}
	T=24.79+27.7n.
	\end{split}
\end{equation}
 
\par For the trimodal distribution in Fig. \ref{Distribution}(b), we estimated the mean transient time $MT_1, MT_2, MT_3$ for each of the three distinct modes,
\begin{eqnarray}\label{transient times }
	MT_1 &=& 467.88 = 24.79 +443.09 \approx + T_1+16 \times T^{'}, \nonumber\\
	MT_2 &= &440.57 = 24.79 +415.78 \approx T_1+ 15 \times T^{'}, \nonumber\\
	MT_3 &=& 413.08 =  24.79+388.29  \approx  T_1+14 \times T^{'}. \nonumber
\end{eqnarray} 
The above expressions support a fact that the variation in total transient time $T$ mainly depends on the number of revolutions during spiraling, which appears as an integral multiple  of $T' (n=14, 15, 16)$ while $T_1$ remains constant ($24.79$) for all the three modes (see Appendix for details). This difference in the number of revolutions leads to three distinct modes of distribution.

\par  For the trajectory in Fig.~\ref{Mech}(d), $T_2\neq0$, and  we find $$T_1=8.39, ~T_2=22.61 ~\mbox{and}~ T^{'}=27.65.$$ For the unimodal distribution in Fig. \ref{Distribution}(e), we estimate the mean transient time for the unique mode as $MT=366.47$ following Eq.~(\ref{tt}), 
$$MT = 366.47 = 8.39 + 22.61 + 335.47 \approx T_1 + T_2 + 12 \times T^{'}.$$ 

\par As we move away from the basin boundary and make a choice of initial points from the subsequent blocks, we see that the trajectories come closer to the $C=0.01$ plane, and in some cases, the trajectory even grazes the plane and spend more time on it as shown in Figs. \ref{Mech}(b-d). As a result, $T_1$ and $T_2$ become larger, and the trajectories spend lesser time during spiraling. This is reflected in the lesser number of revolutions undertaken by the trajectory before reaching the predator-free state, and hence less number of modes are seen. Furthermore, in Fig.~\ref{Mech}(d), the trajectory takes a large amount of time on the $C=0.01$ plane and shows very few revolutions during spiraling, rather quickly arriving at the predator-free limit cycle.

\par The fact of decreasing mean transient time ($MTT$) with distance from the basin boundary as shown in Fig. \ref{Distribution}(f), is also  explained from Figs.~\ref{Mech}(a-d). As the initial point is chosen from a location far away from the basin boundary, a larger transient time is spent by the trajectories on the $C=0.01$ plane. Since this plane makes the trajectories approach the predator-free period-1 limit cycle faster than approaching it through spiraling motion in the space without touching the vertical yellow plane, the mean transient time decreases as the initial points are chosen further away from the basin boundary. The transient time and standard deviation (fluctuations) for different modes for all the colored boxes are presented in Table \ref{T2}. \FloatBarrier
\begin{table}
	\captionof{table}{Statistical measures} \label{T2}
	\begin{tabular}{| c | c | c | c | c | } 
		\hline 
		Color & Mode Sl. no. & MTT & MTT (all modes) & Std. deviation \\
		\hline \hline
		Green & 1 & 413.08 & & 2.07 \\
		& 2 & 440.57& 440.51 & 3.08 \\  
		& 3 & 467.88& & 1.39 \\
		\hline
		Magenta & 1 & 390.75 & 401.15 & 1.31\\
		& 2 & 411.54 & & 0.72\\
		\hline
		Black & 1 & 370.58 & 379.34 & 0.52\\
		& 2 & 388.10 & & 0.67\\
		\hline
		Red & 1 & 366.47& 366.47 & 0.97\\
		\hline		    
	\end{tabular}
\end{table}

\section{Basin Structure: Distribution pattern of transients}\label{basin sturcture}
\par
\begin{figure}[ht!]
	\hspace*{-4cm}
	\begin{subfigure}[h!]{0.2\textwidth}
		\centering
		\subcaptionbox{}
		{
			\includegraphics[width=2\textwidth,height=5.7cm]{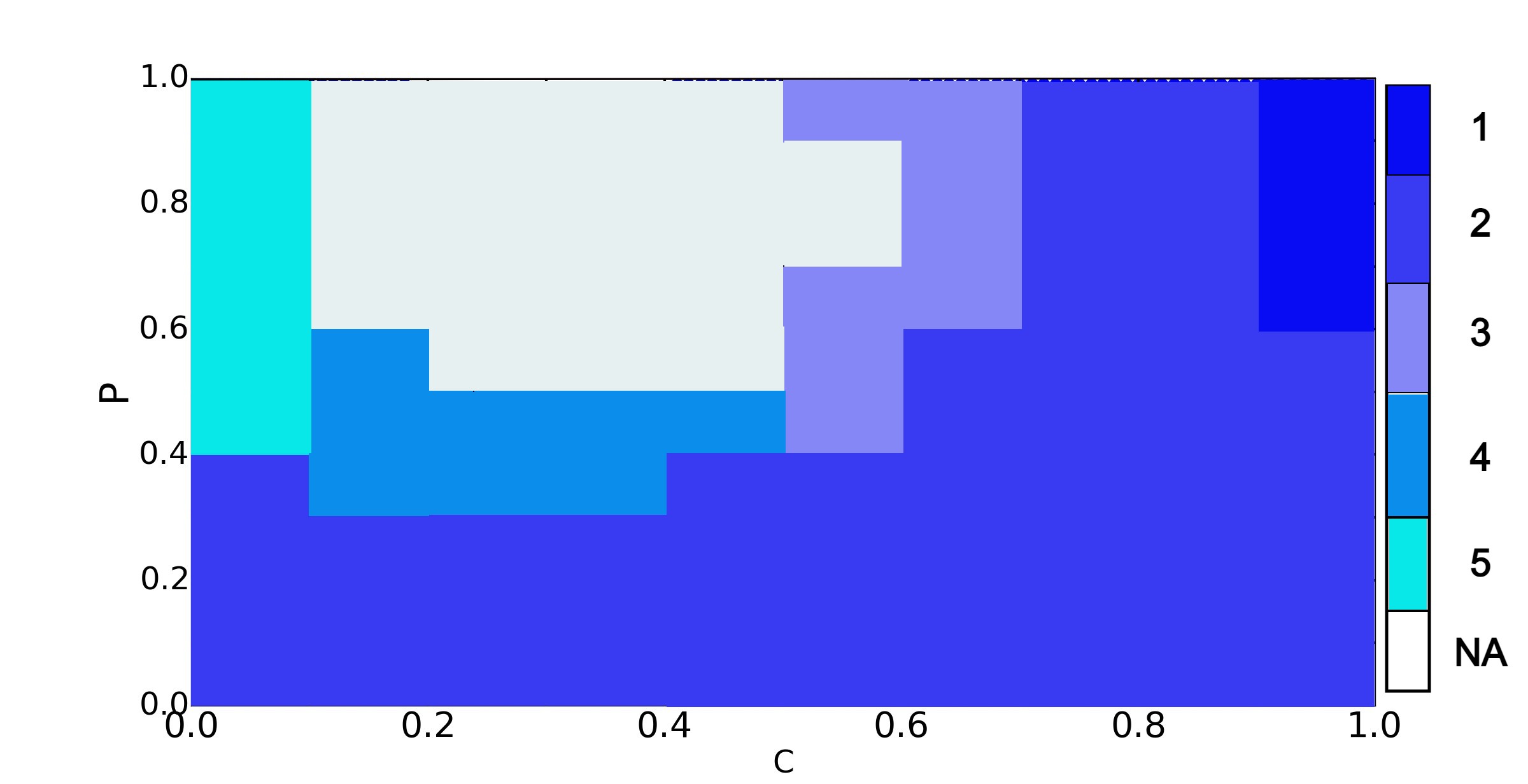} %
		}\qquad		
	\end{subfigure}\\
	\hspace*{-4cm}
	\begin{subfigure}[h!]{0.2\textwidth}
		\centering
		\subcaptionbox{} 
		{
			\includegraphics[width=2\textwidth,height=5.7cm]{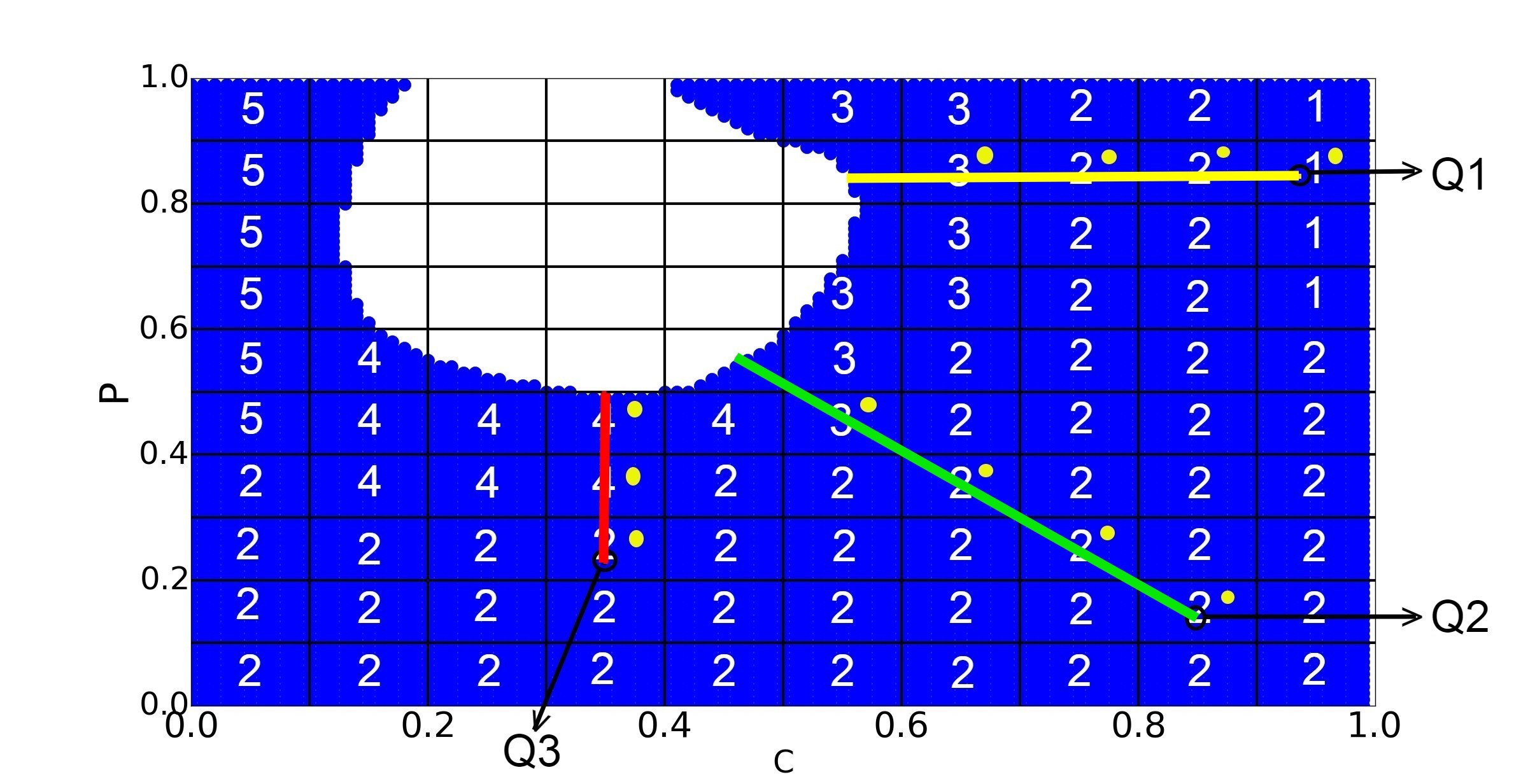}%
		}\qquad	
	\end{subfigure} 
	\caption{{\bf Pattern of distribution of transients in the basin of the tri-tropic food chain model.} (a) Various color shades in the basin correspond to the number of modes in the distribution of transient time for each $N_i$. The color bar denotes the number of modes. (b) Numbers in each block mark the number of modes in the distribution of transient time. Three lines (yellow, green, and red) represent the shortest distance from the basin boundary to the points ($Q_1, Q_2, Q_3$), where a homogeneity index is calculated. Yellow circles mark the boxes through which the lines pass.}
	\label{Heterogeneity}		
\end{figure}

 $\;$  $\;$ In order to find a broader picture of the distribution pattern of transient time throughout the basin, we divide the entire blue region (Fig. \ref{Distribution}a) into identical square blocks $N_{i}, i \in \mathbb{Z_+}$, of edge length $10^{-1}$ units such that 
$N_{i}=\{(x, y)\in \mathbb{R_+}^2: 0.1(n-1) < x \leq 0.1n,~ 0.1(m-1)< y \leq 0.1m,~~ n, m = 1, 2, ...., 10\}.$ From each $N_{i}$, a number of $10^5$ random initial points are chosen, and the distribution of transient time of the trajectories is mapped in Fig. \ref{Heterogeneity}(a), where the number of modes in a square block $N_i$, $M(N_i)$, is represented by the color bar. Keeping at par with the previous section, the number of modes is found larger in the region near the boundary between the white and blue basins. As we move away from the basin boundary, the number of modes decreases and becomes minimal at the farthest location from the basin boundary. The basin clearly shows fluctuation in the number of modes that vary with the location of initial points. More categorically, as shown in Fig.~\ref{Heterogeneity}(b), a 5-mode distribution is seen on the left-hand side of the white basin. The bottom and the right-hand side of the boundary displays $4,~ 3,~ 2$-mode distribution with a unimodal distribution at top right corner of the basin. The basin is mainly dominated by bimodal distributions (dark blue), with traces of unimodal distributions existing at the upper right corner. Besides such pattern in the distribution of modes, directional variation in the number of modes exists, indicating an anisotropy in the distribution pattern of transients in the basin. We quantify such inhomogeneity and anisotropy of the basin structure by introducing two new metrics.

\subsection{Inhomogeneity of distribution pattern of transients in the basin}
\par $\;$  $\;$ The very presence of multimodal distributions in the transient time, obtained from initial points in the blue basin, implies inhomogeneity of the basin topology \cite{kempf2004encyclopedia}. To quantify this property, we define a homogeneity index ($HI$) of any point $Q$ in the basin. We consider the shortest distance, $L$, measured from the point $Q$ to the boundary of the white basin, and calculate the reciprocal of the mean of the number of modes, $\{M(N_i)\}$, in all the square blocks, $\{N_i\}$, through which the $L$ line passes. Thus the homogeneity index of the point $Q$, $HI(Q)$, is given by
$$HI(Q) = \left[\frac{\Sigma(M(N_i))}{\mathbf{card}(N_i)}\right]^{-1}$$,
where $\mathbf{card}(N_i)$ is the number of boxes through which the line $L$ passes and $M(N_i)$ is the number of modes in the block $N_i$.\\
For illustration, we consider three points $Q_1,\,Q_2,\,Q_3$ in the blue basin in Fig.~\ref{Heterogeneity}(b). The shortest distance for each point $Q_i$, ($i=1,2,3$) from the basin boundary is demarcated by three lines (yellow, green, and red lines, respectively). By our definition, the homogeneity index of each point is then given by
\begin{eqnarray}\label{ homogeneity index}
	HI(Q_1) &=& \left[\frac{3+2+2+1}{4}\right]^{-1} = 2^{-1}=0.5 \nonumber\\
	HI(Q_2) &=& \left[\frac{3+2+2+2}{4}\right]^{-1} = 2.25^{-1}=0.44\nonumber\\
	HI(Q_3) &=& \left[\frac{4+4+2}{3}\right]^{-1} = \left[\frac{10}{3}\right]^{-1}=0.3\nonumber
	\end{eqnarray}
	
If all the squares boxes in a basin produce unimodal distributions, then $HI(N_i)=1\; \forall \; i$, and it is said to be a homogeneous basin. On the other hand, if $HI(N_i)<1$, the basin is inhomogeneous as it is found true for the basin of our present study.

\par One can easily verify that the following two results are true for $HI$:
\begin{itemize}
	\item (i)  $0\leq$ $HI$ $\leq 1$, and
	\item(ii) along any fixed direction, $HI$ is a non-decreasing function of the distance from the white basin.
\end{itemize}
 It is obvious that $HI\geq 0$. To show that $HI$ of any point $Q$ cannot exceed $1$, we note that
\begin{eqnarray}
HI(Q)=\left[\frac{\Sigma(M(N_i))}{\mathbf{card}(N_i)}\right]^{-1} &=& \frac{\mathbf{card}(N_i)}{\Sigma(M(N_i))} \nonumber \\
&\leq& \frac{\mathbf{card}(N_i)}{\mathbf{card}(N_i)} ~~\nonumber \\
&=&1\nonumber 
\end{eqnarray}
$[\because \mathbf{card}(N_i)\leq  \Sigma(M(N_i)),~ since ~ M(N_i)\geq 1]$\\
To show the second result, we consider a line $L$, whose one end is on the basin boundary. Now take any two points $G_1,\,G_2$ on $L$ such that $G_2$ is at a greater distance from the white basin than $G_1$. For the sake of simplicity, we consider the reciprocal of the homogeneity index $HI^{-1}$, which is essentially the arithmetic mean of the number of modes, $\{M(N_i)\}$, where $\{N_i\}$ are boxes through which $L$ passes to reach the desired point. It becomes clear from Fig. \ref{Heterogeneity}(a), that for a particular box $N_i$, $M(N_i)$ remains constant or decreases as $N_i$ gets further away from the basin boundary. Thus the sequence $\{M(N_i)\}$, along any line, becomes a non-increasing sequence. Since we know that the sequence of boxes leading to $G_1$ through $L$, is a subsequence of the sequence leading to $G_2$, we infer $\{N_i\}|_{G_1}\subseteq\{N_i\}|_{G_2}$ and hence $\{M(N_i)\}|_{G_1}\subseteq\{M(N_i)\}|_{G_2}$. Now from the fact that the arithmetic mean of a monotonically non-increasing sequence is also non-increasing \cite{little2015real}, we conclude 
$$HI(G_1)^{-1} \geq HI(G_2)^{-1} \implies HI(G_2) \geq HI(G_1).$$
Hence $HI$ is non-decreasing.
\subsection{Anisotropy of distribution pattern of transients in the basin}
\par $\;$  $\;$ Any pattern is called isotropic if the same properties are maintained  at equal distances in all directions from any fixed point of a space \cite{little2015real}. As observed in Fig.~\ref{Heterogeneity}(b), after traversing four square blocks along the yellow line unto $Q1$, we arrive at a box of unimodal distribution of transient time. On the other hand, traveling along the green line and covering an equal number of square blocks, a bimodal distribution is encountered at $Q2$. This indicates a directional dependence of the number of modes of distribution in the blue basin, which appears as anisotropic. We attempt a quantification of this structural property of the basin from the point of view of transient time.

\begin{figure}[ht!]
	\hspace*{-4cm}
	\begin{subfigure}[h!]{0.2\textwidth}
		\centering
		\subcaptionbox{}
		{
			\includegraphics[width=2\textwidth,height=5.7cm]{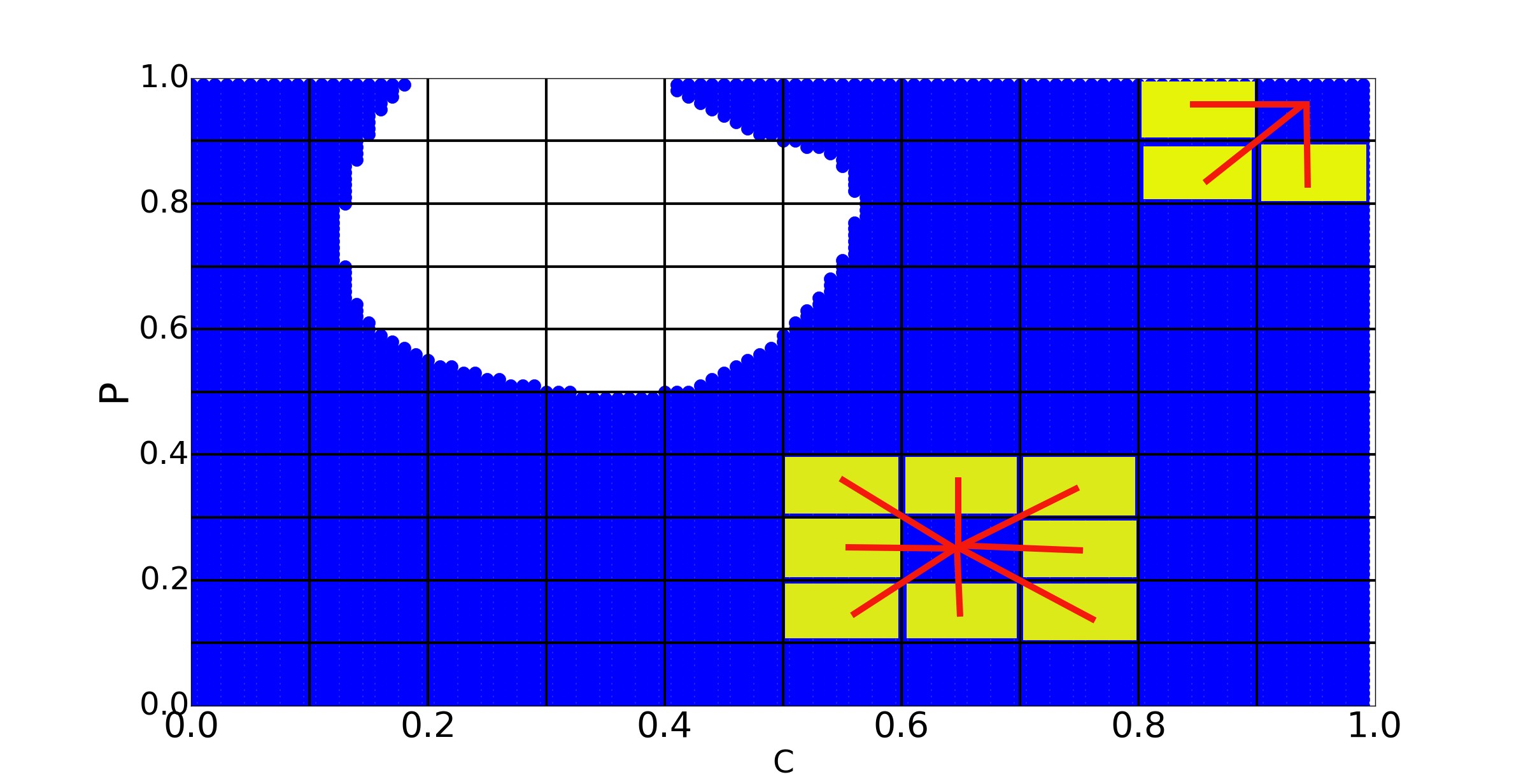} %
		}\qquad		
	\end{subfigure}\\
	\hspace*{-4cm}
	\begin{subfigure}[h!]{0.2\textwidth}
		\centering
		\subcaptionbox{} 
		{
			\includegraphics[width=2\textwidth,height=5.7cm]{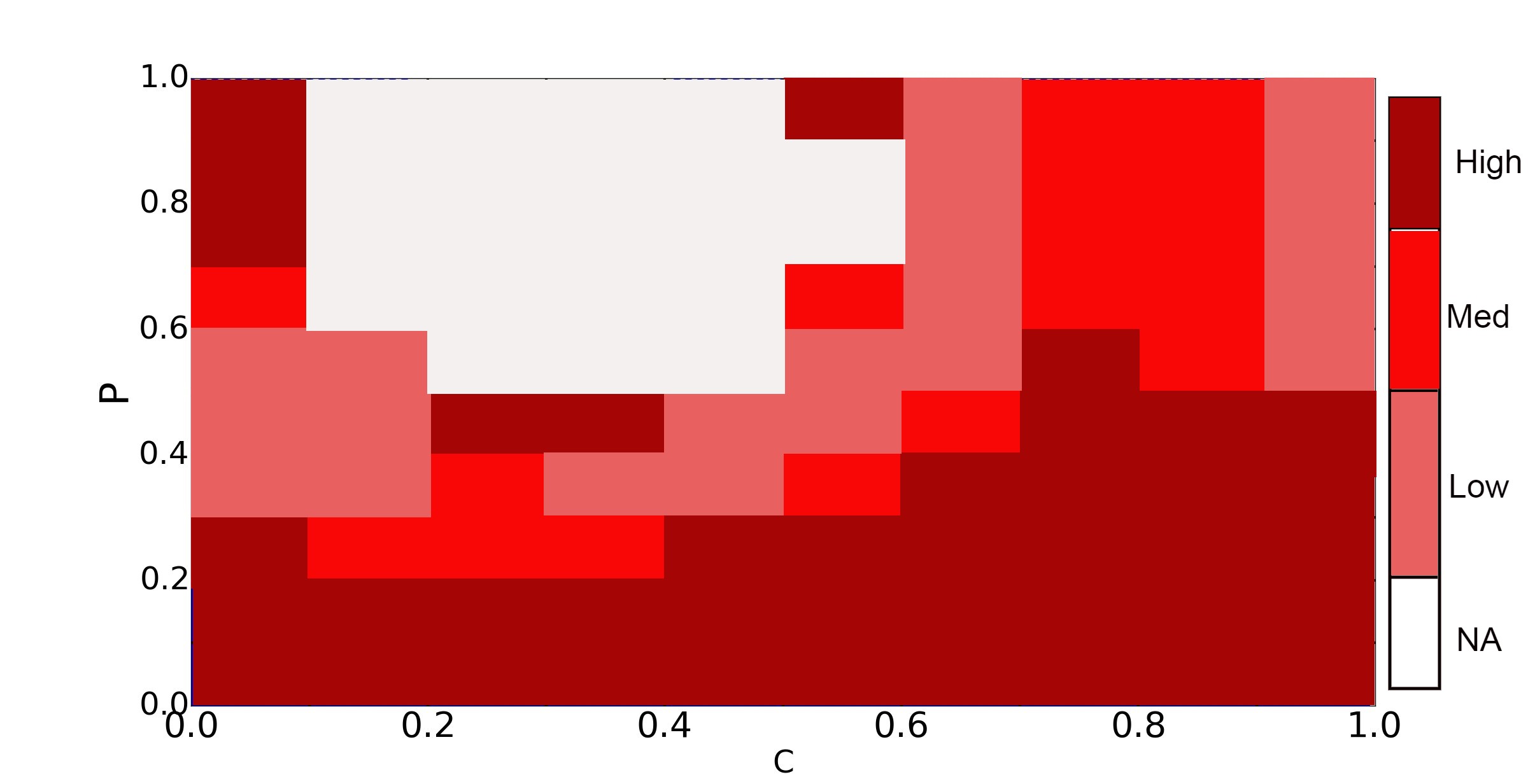}%
		}\qquad	
	\end{subfigure} 
	\caption{{\bf Anisotropic pattern of distribution of transients in the basin.} (a) Two blue boxes are shown surrounded by immediate neighbors of yellow square blocks called box neighborhoods. Red lines indicate the paths through which the yellow regions could be reached from the blue boxes without passing through any other square. (b) Map of the local isotropic index ($LII$). Dark red (high locally isotropic) $0.8 \leq LII \leq 1.0$, Red (moderate locally isotropic) $0.6 \leq LII < 0.8$, Light red (poor locally isotropic) $LII <0.6$ (white color is not considered for calculation).}
	\label{Noniso}		
\end{figure}

\par Before quantifying this isotropic property of a basin, we first define a box neighborhood ($BN$). Let us consider a square block $N_i$, then the box neighborhood, $BN(N_i)$, of $N_i$ is defined as the square blocks $N_j$ $(j \neq i)$ such that $N_j$ can be accessed from $N_i$ without crossing another box. In Fig.~ \ref{Noniso}(a), the yellow squares around the two boxes at the locations $[0.6,0.7]\times[0.2,0.3]$ and $[0.9,1.0]\times[0.9,1.0]$ represent their box neighborhood. 

\par Now we define a local isotropic index ($LII$) of a box $N_i$ as the probability of finding square blocks in a box neighborhood of $N_i$ such that they have the equal number of modes in their transient time distribution as $N_i$. Thus, a $LII$ of the box $N_i$ is defined by
\begin{equation}
\begin{split}
LII(N_i)=\frac{\mathbf{card}\{N_j: j \neq i, N_j \in BN(N_i), M(N_j)=M(N_i)\}}{\mathbf{card}(BN(N_i))},
\end{split}
\end{equation}
where $\mathbf{card}(X)$ is the cardinality of the set $X$.

\par For illustration purposes, we calculate the $LII$ of the two concerned boxes in Fig. \ref{Noniso}(a), namely $[0.6,0.7]\times[0.2,0.3]$ $(N_1)$ and $[0.9,1.0]\times[0.9,1.0]$
$(N_2)$. We see from Fig. \ref{Heterogeneity}(a), $M(N_1)=2$ and $M(N_2)=1$ and from Fig. \ref{Noniso}(a), $BN(N_1)=8$ and $BN(N_2)=3$. Furthermore from Fig. \ref{Heterogeneity}(a), the number of boxes with same number of modes as $N_1$ and $N_2$ in their respective box neighborhoods are 8 and 1 respectively. Hence by definition, 
\begin{eqnarray}
LII(N_1) &=& \left[{\frac{8}{8}}\right] = 1 \nonumber\\
LII(N_2) &=& \left[\frac{1}{3}\right] = 0.33 \nonumber
\end{eqnarray}
We now calculate the $LII$ of each $N_i$ and categorize the entire basin into four groups listed in Table \ref{T3}.
\begin{table}
	\centering
	\captionof{table}{Local isotropic indices} \label{T3}
	\begin{tabular}{| c | c| c | } 
		\hline 
		Color & Description & Range\\
		\hline \hline
		Dark red & highly locally isotropic & 0.8-1.0 \\
		Red & moderately locally isotropic & 0.6-0.8 \\  
		Light red & poorly locally isotropic & 0.0-0.6 \\
		White & region not considered & -\\
		\hline		    
	\end{tabular}
\end{table}

\begin{figure}[h!]
	\hspace*{-4cm}
	\begin{subfigure}[h!]{0.2\textwidth}
		\centering
		\subcaptionbox{}
		{
			\includegraphics[width=2\textwidth,height=5.7cm]{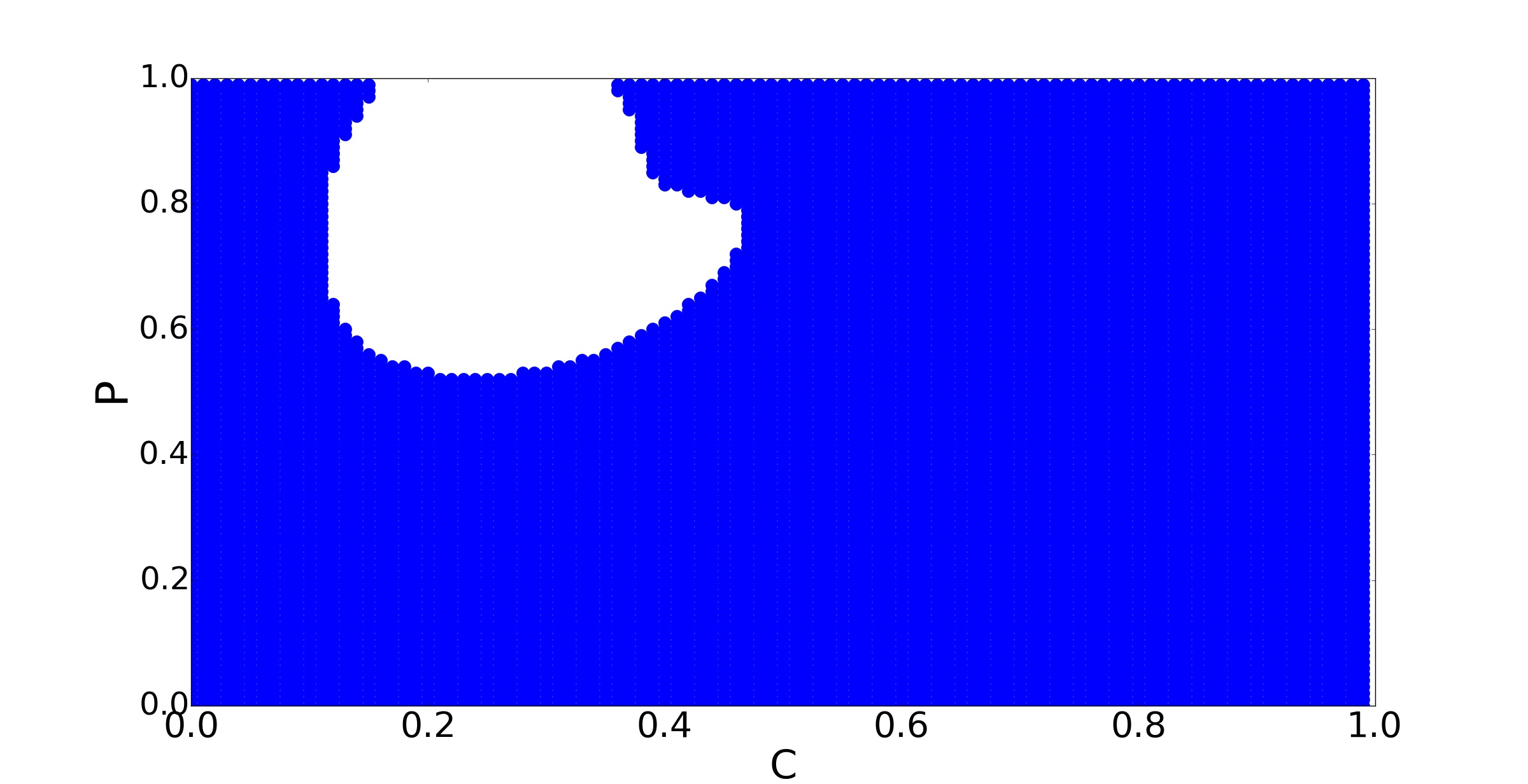} %
		}\qquad		
	\end{subfigure}\\
	\hspace*{-4cm}
	\begin{subfigure}[h!]{0.2\textwidth}
		\centering
		\subcaptionbox{} 
		{
			\includegraphics[width=2\textwidth,height=5.7cm]{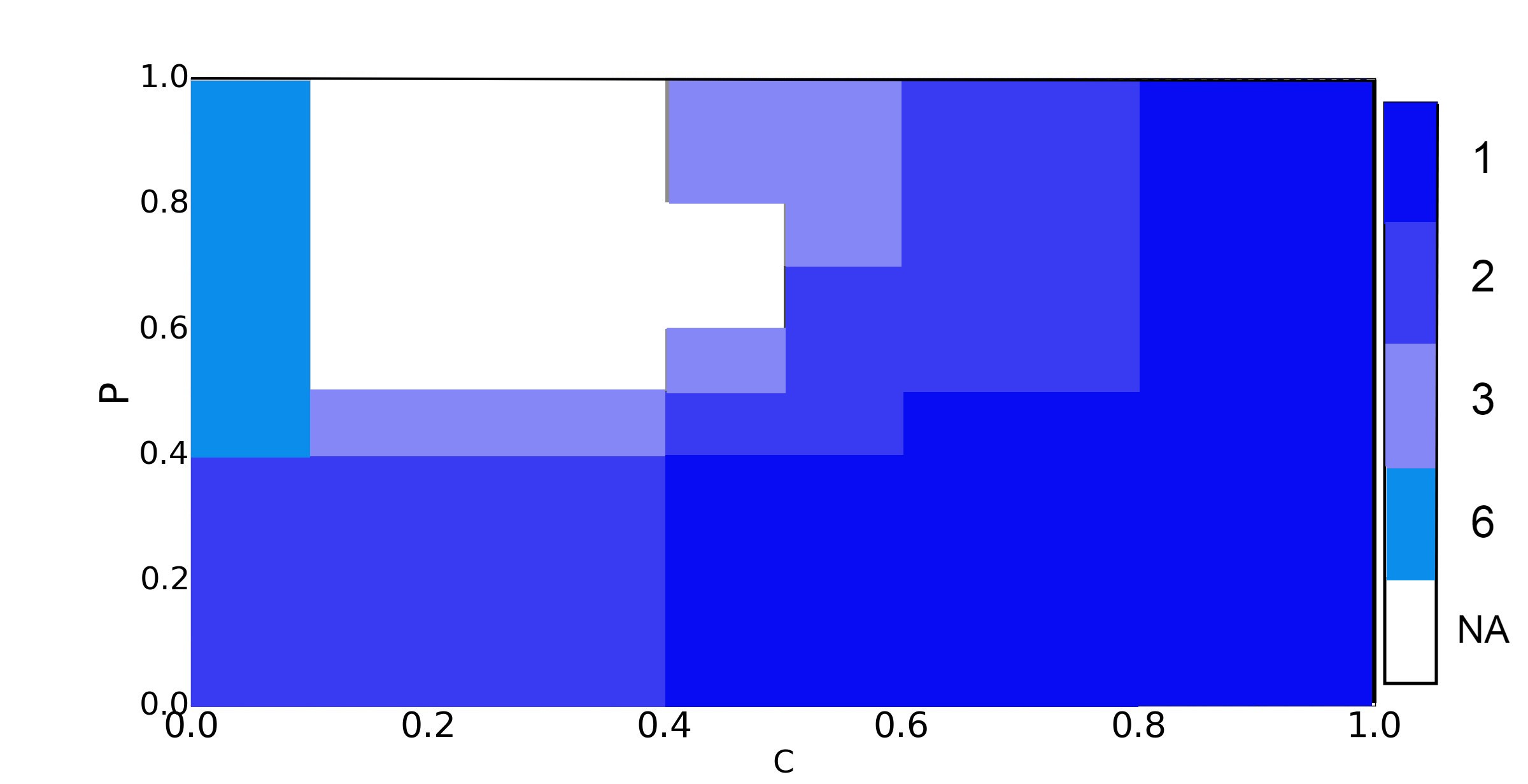}%
		}\qquad	
	\end{subfigure}\\
	\hspace*{-4cm}
	\begin{subfigure}[h!]{0.2\textwidth}
		\centering
		\subcaptionbox{}
		{
			\includegraphics[width=2\textwidth,height=5.7cm]{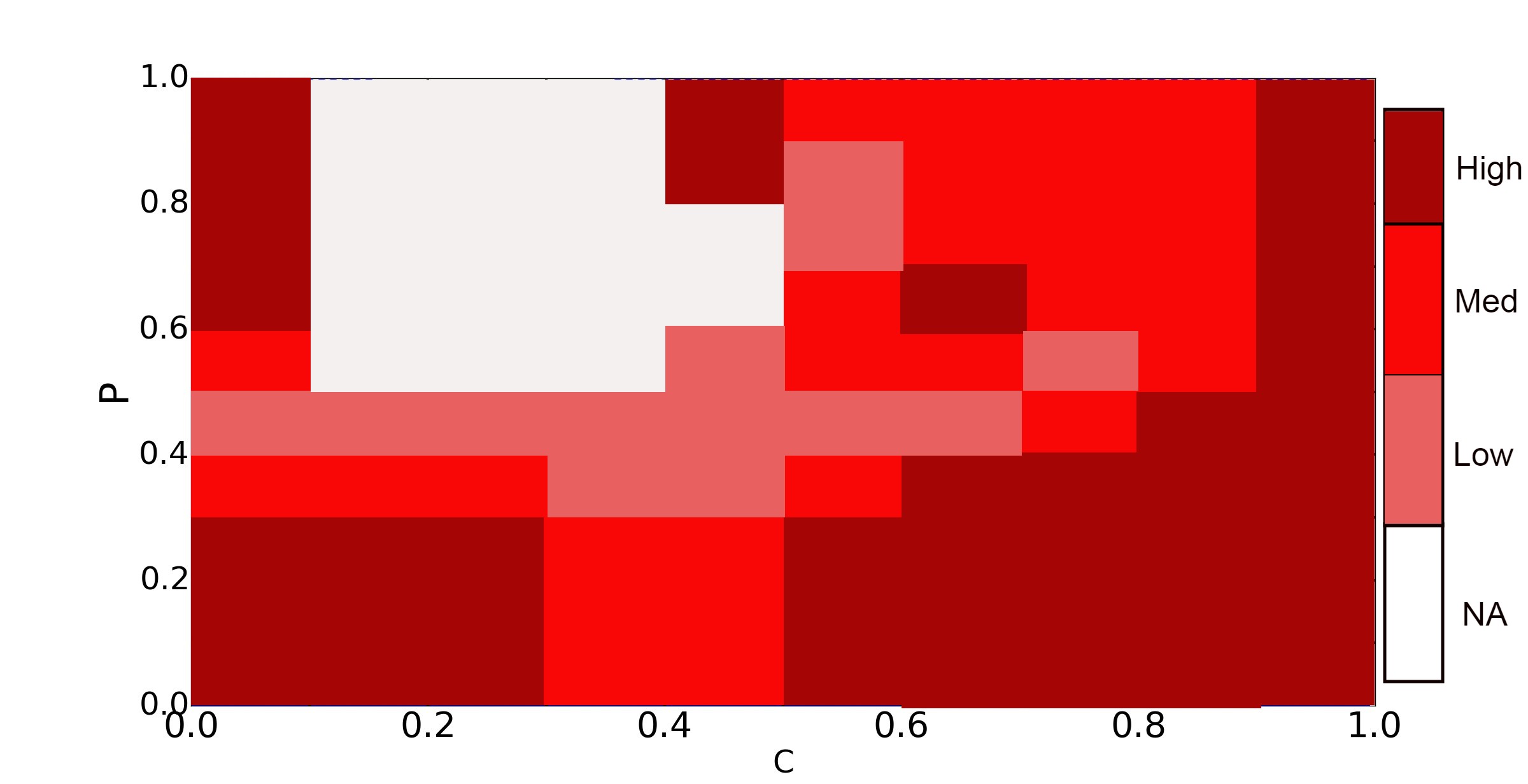} %
		}\qquad		
	\end{subfigure} 
	\caption{{\bf Homogeneity index ($HI$) and local isotropic index ($LII$) against initial resource level $R=0.9$.} (a) Basin of attraction taking $R=0.9$. (b) Variation in $HI$. Number of modes (indicated by a color bar) in each square region when the basin is plotted on the plane $R=0.9$. (c) Distribution of $LII$. Colors carry the same meaning as in Table \ref{T3}.}
	\label{R9}		
\end{figure}

\subsection{Effect of resource enrichment}
\par $\;$ $\;$ We check the variation in the distribution of the number of modes in the basin with enrichment of resources with a different value of $R=0.9$ instead of $R=0.5$ and plot $HI$ and $LII$ in Fig.~\ref{R9}. Quite understandably, the inhomogeneity pattern changes with resource enrichment. Figure~ \ref{R9}(a) shows that the unimodal distribution of the transient time is a more dominant feature instead of the dominant bimodal scenario in Fig.~\ref{Heterogeneity}(a). However, on the left-hand side of the white basin, a 6-mode distribution pattern is seen while for the rest of the region, $M(N_i)\mid_{R=0.5} \geq M(N_i)\mid_{R=0.9}$. It depicts that the same box in Fig. \ref{Heterogeneity}(a) produces a distribution with a higher or equal number of modes than it does when the initial resource population value is increased. This results in an increment of the average values of $HI$. 

\par On the contrary, we see no significant changes between Fig. \ref{Noniso}(b) and Fig. \ref{R9}(b). Since changing the resource plane from $R=0.5$ to $R=0.9$ only alters the shape of the basins, $LII$ being a property independent of the location and geometry of the basin boundary and only pertaining to the local box neighborhood of the boxes ($N_i$), does not show any noticeable change.

\section{Conclusion}\label{discussion}
\par $\;$ $\;$ In a classic resource-consumer-predator bistable food chain, the predator shows extinction after an elapse of time called as transients when a state of all coexisting species goes to a predator-free state after a transient time. The distribution of the transients of predator extinction appears as distinctively multimodal with a local variation in the number of modes that appears as inhomogeneous and anisotropic relative to a basin boundary separating two basins of the coexisting states. To be precise, the number of modes as well as the mean transient time decreases with increase in the distance from the basin boundary. We explained the reason behind the existence of such distinct modes and the local variation in the number of modes and mean transient time in the basin of a predator-free state from the dynamical nature of the trajectories. We have quantified the characteristic structural properties of the basin, in terms of the number of modes in distribution by introducing two new metrics viz., homogeneity index and local isotropic index. We witnessed an increasing homogeneity in the basin when the system is enriched with resources. 

\par The fate of the food chain is strongly influenced by the initial species population. The food chain can either witness coexistence of all the species or a state predator-free state after extinction. While considering the basin of initial population densities that drive the top predator to extinction, we observe existence of an intermediate state where all the three species coexist, before the predator density finally becomes zero. The ecological implications of such transient dynamics of species population and its duration, are profound. The duration of existence of the intermediate state, called transient time, is seen to have a clustering effect on the initial population densities, where they are apparently divided into various subpopulations, as evident from the multimodal distributions of transient time \cite{kempf2004encyclopedia}. Each subpopulation or group is seen to generate transient times equal to or close to a particular principal value or mode. The entire set of initial population densities, triggering predator extinction, can be divided into various subsets or groups of suitable resolution. This division is characterized here by two properties, inhomogeneity and anisotropy. From the trend of inhomogeneity, we see that the initial population densities are divided into fewer groups, as their difference with the critical initial values, which decides two different fates of the ecosystem, increases. The anisotropic factor reveals that the number of groups of species subpopulations may change even if the difference between the critical and pertaining initial population densities remains the same. Interestingly, when we enrich the initial resource population in the food chain, the initial populations densities become more homogeneous as reflected from the uniform distribution of transients and are thus divided into fewer subgroups.

\par Extinction of species at any trophic level may cause the loss of energy flow to the higher trophic levels, leading to a trophic cascade. Thus, species persistence in a food structure is vital to prevent regime shift and ecosystem functioning. Our analysis reveals that mean transient time is an increasing multi-valued function of the number of modes in distribution. Broadly speaking, the initial values generating multimodal distributions produce trajectories with longer transient time than the ones producing unimodal distributions. It tells us that  there exists a wider {time window} for protecting the predators from extinction using suitable restoration mechanisms. Our initial study (see Appendix)  indicates that the probability of success of any restoration mechanisms is higher and the associated cost is lower for the location of initial states  leading to multimodal distribution than in the unimodal one. A rigorous study is needed to ascertain which restoration mechanisms is appropriate and cost-effective to save the extinction of species with a higher probability. This is an open problem for future study. \\

\noindent {\bf Acknowledgment:} Research is supported by RUSA 2.0, Jadavpur University, Ref No. R-11/725/19, dated 26.06.2019. A.M. is supported by the National Science Centre, Poland, OPUS Programme Project No. 2018/29/B/ST8/00457. \\

\noindent{\bf Data availability}: All data and numerical codes are available from the authors free on request.\\
\section*{Appendix}
\subsection{Calculation of average spiraling time}
\begin{figure*}[ht!]
	\hspace*{-4cm}
	\begin{subfigure}[h!]{0.2\textwidth}
		\centering
		\subcaptionbox{}
		{
			\includegraphics[width=2\textwidth,height=5.7cm]{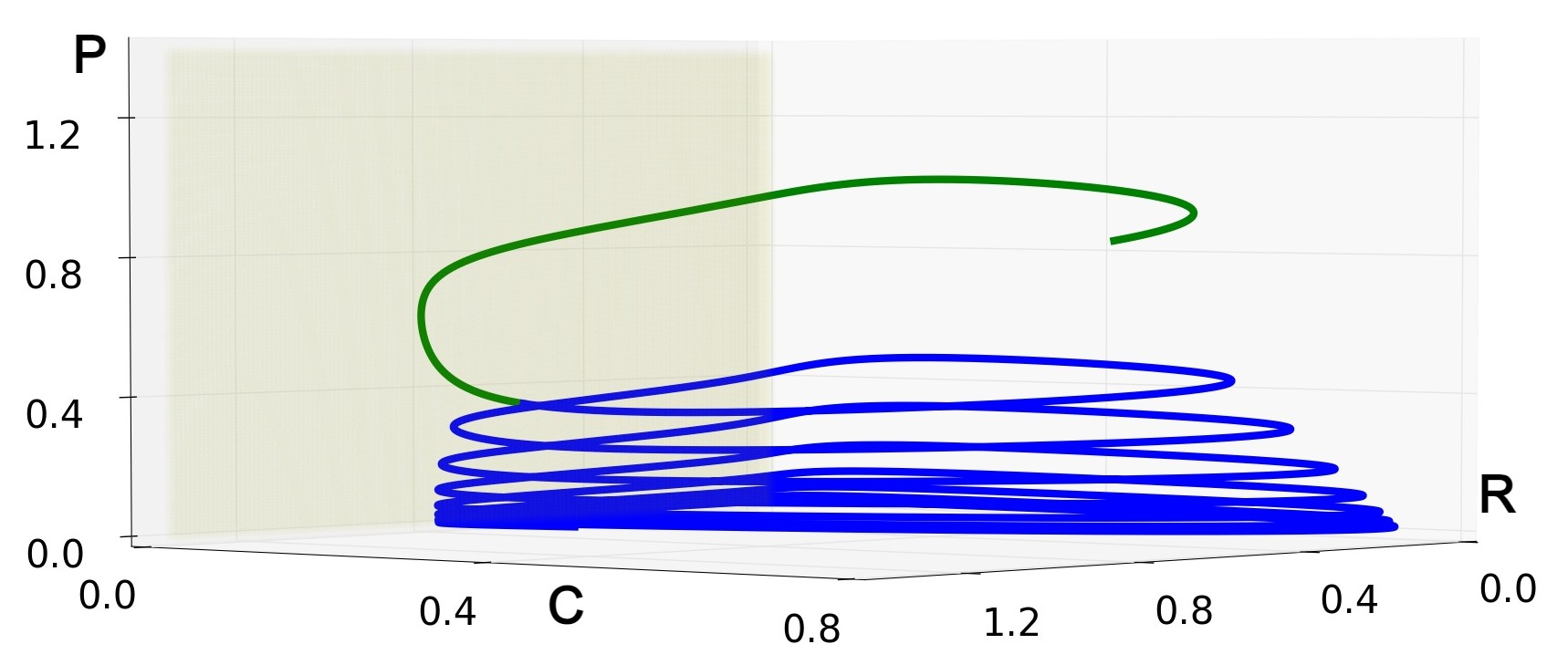} %
		}\qquad		
	\end{subfigure}\hspace*{3.5cm}
	\begin{subfigure}[h!]{0.2\textwidth}
		\centering
		\subcaptionbox{} 
		{
			\includegraphics[width=2\textwidth,height=5.7cm]{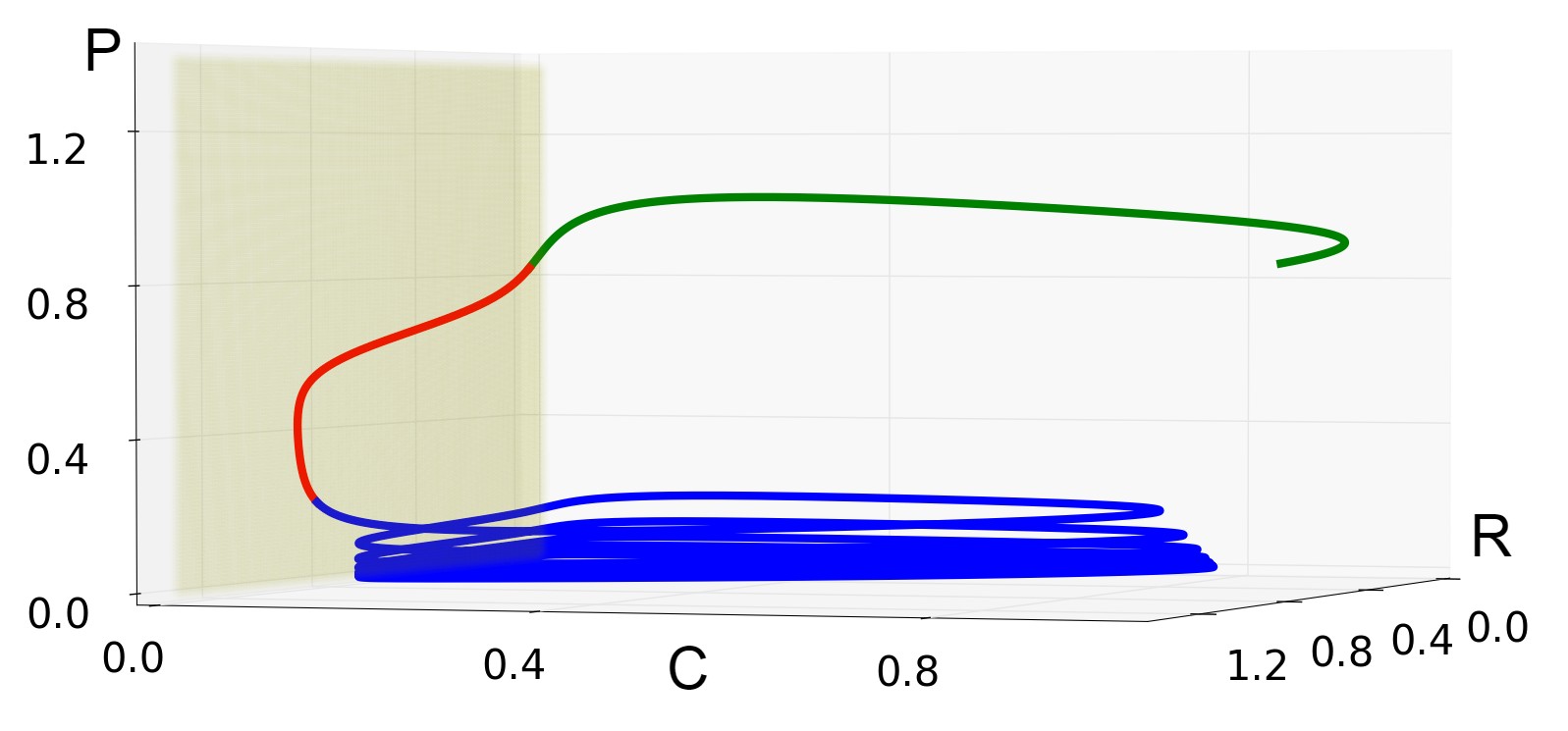}%
		}\qquad	
	\end{subfigure}\\
	\hspace*{-4cm}
	\begin{subfigure}[h!]{0.2\textwidth}
		\centering
		\subcaptionbox{}
		{
			\includegraphics[width=2\textwidth,height=5.7cm]{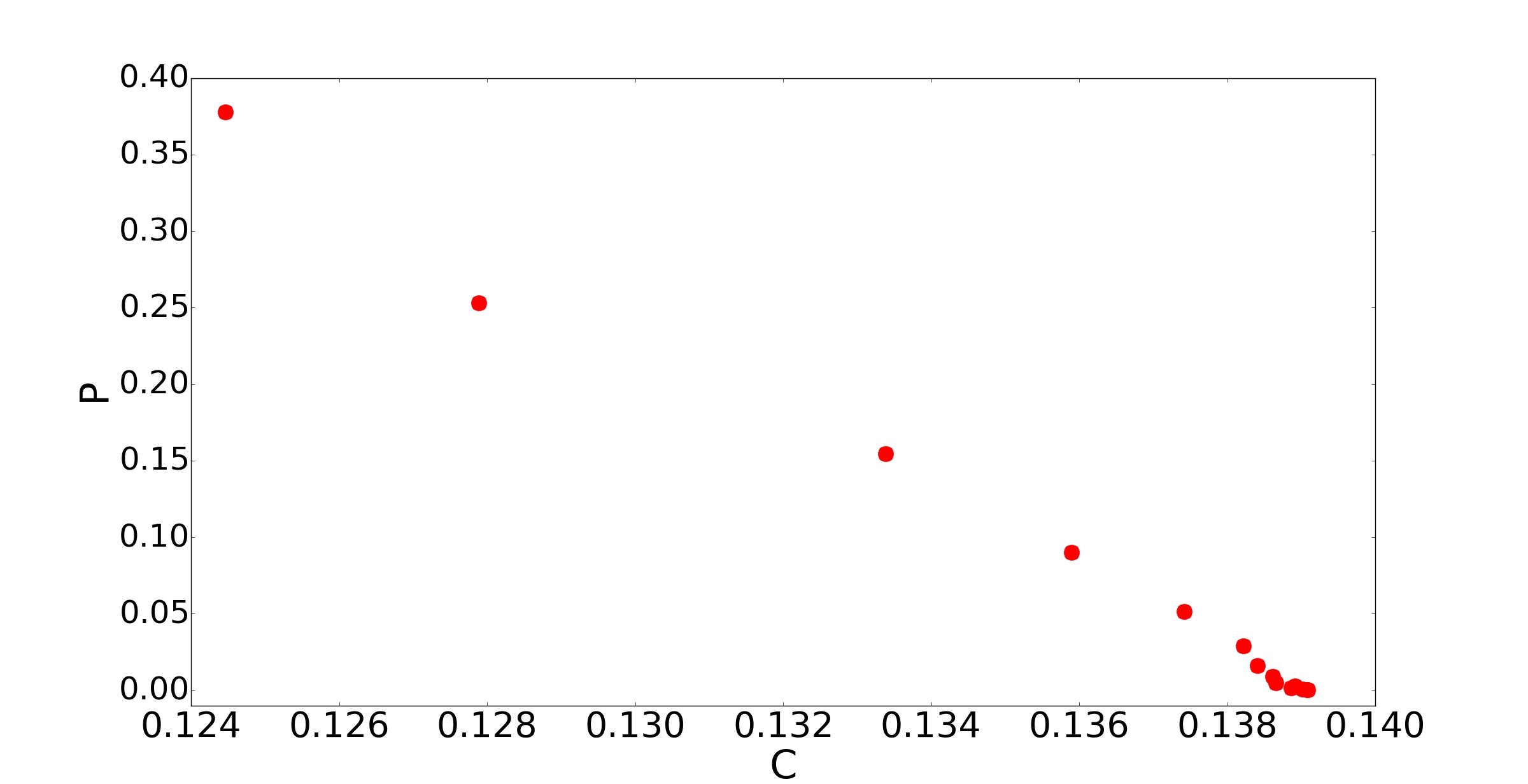} %
		}\qquad		
	\end{subfigure}\hspace*{3.5cm}
	\begin{subfigure}[h!]{0.2\textwidth}
		\centering
		\subcaptionbox{} 
		{
			\includegraphics[width=2\textwidth,height=5.7cm]{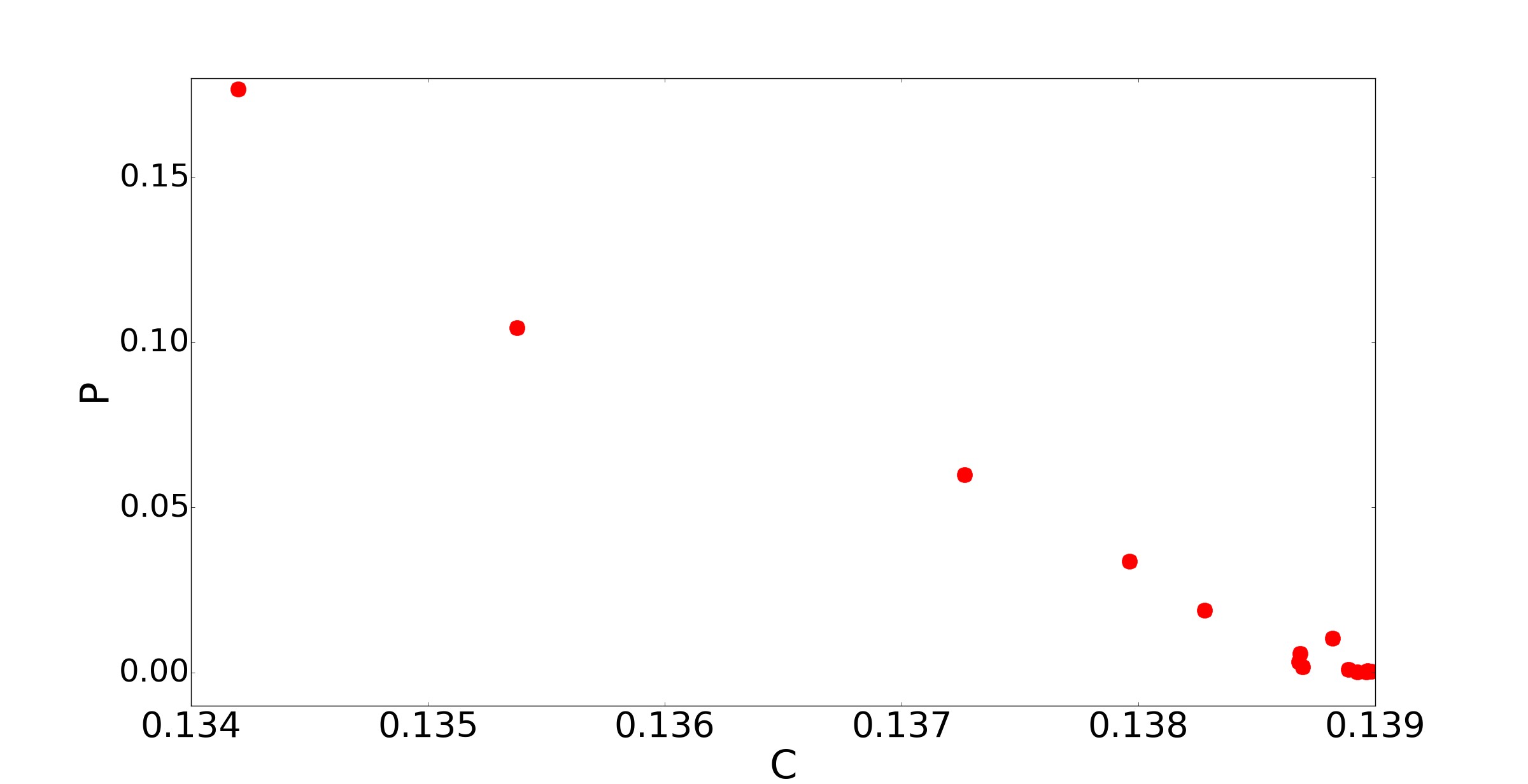}%
		}\qquad	
	\end{subfigure}
	\caption{Initial points are (0.5,0.65,0.85) for (a) and (0.5,0.95,0.85) for (b). (a,b) Green curve corresponds to $T_1$, red curve to $T_2$ and blue curve to $T_3=nT^{'}$. (c,d)  Poincar\'e return map of (a) and (b) respectively, taking $R=0.87$ as the Poincar\'e plane.  Parameters are chosen from Table \ref{T1}.}
	\label{calc}		
\end{figure*}

\par For numerical simulations, we have used the $4^{th}$ order Runge-Kutta algorithm, taking temporal step size as 0.01. We elaborate here the methodology for calculating the metrics required to quantify the different segments ($T_1,$ $T_2$ and $T^{'}$) of the total transient time ($T$) for the Fig. \ref{Mech}(a,d) (see Section \ref{distribution of transient time}). The calculation of the same for other figures follows a similar manner.

\par The numerical solution of Eq. (\ref{Model}) is plotted in Fig. \ref{calc}(a), such that the green line represents the path of the trajectory from its initial point (0.5,0.65,0.85) to the phase where it starts spiraling. On the other hand, the blue line represents the spiraling phase of the trajectory. For calculating the time required to reach the spiraling phase ($T_1$) and the subsequent mean revolution time ($T^{'}$), we consider the Poincar\'e return map of Fig. \ref{calc}(a), by taking the Poincar\'e plane as $R=0.87$. The plane is chosen so that it cuts the curve almost at the point where the trajectory enters the spiraling motion, demarcated by the change in the color of the trajectory from green to blue. Thus, the time corresponding to the left-most dot in Fig. \ref{calc}(c) represents the end of the green curve of Fig. \ref{calc}(a), that is $T_1$. The considered dots correspond to the points where the curve cuts the Poincar\'e plane after each spiral revolution and hence the time lapse between each dot gives us the corresponding revolution time. For figuring out $T^{'}$, we note the times, say $F_i, ~i=1,2,3,....n$, corresponding to each dot (red dots in Fig. \ref{calc}(c)), where $n$ is the total number of dots in Fig. \ref{calc}(b). Observe that $n$ is $14$ here. Therefore, the mean of the differences between the times corresponding to the adjacent dots gives the mean revolution time, $T^{'}$. Thus, 
$$T^{'}=\frac{\sum_{i=1}^{n-1}(F_{i+1}-F_i)}{n-1}.$$

\par Fig. \ref{calc}(b) is plotted in the same way as Fig. \ref{calc}(a), but with initial points (0.5,0.95,0.85), such that for this case $T_2\neq0$. The calculation of $T^{'}$ from the Poincar\'e return map (Fig. \ref{calc}(d)) is done in the same manner as the previous case. For the determination of $T_1$ and $T_2$, we simply note the time when the green curve of Fig. \ref{calc}(b) hits the light yellow plane $C=0.01$, thus obtaining $T_1$, and also note the same when the red curve exists the yellow plane. The difference in the time of exiting and entering the plane $C=0.01$ gives us $T_2$.
  
\par The magnitude of each metric concerning the calculation of transient time is calculated by taking the mean of the results obtained after 50 simulations, choosing random initial points from the pertaining colored box.

\begin{figure*}[ht!]
	\hspace*{-5cm}
	\begin{subfigure}[h!]{0.2\textwidth}
		\centering
		\subcaptionbox{}
		{
			\includegraphics[width=2.5\textwidth,height=5cm]{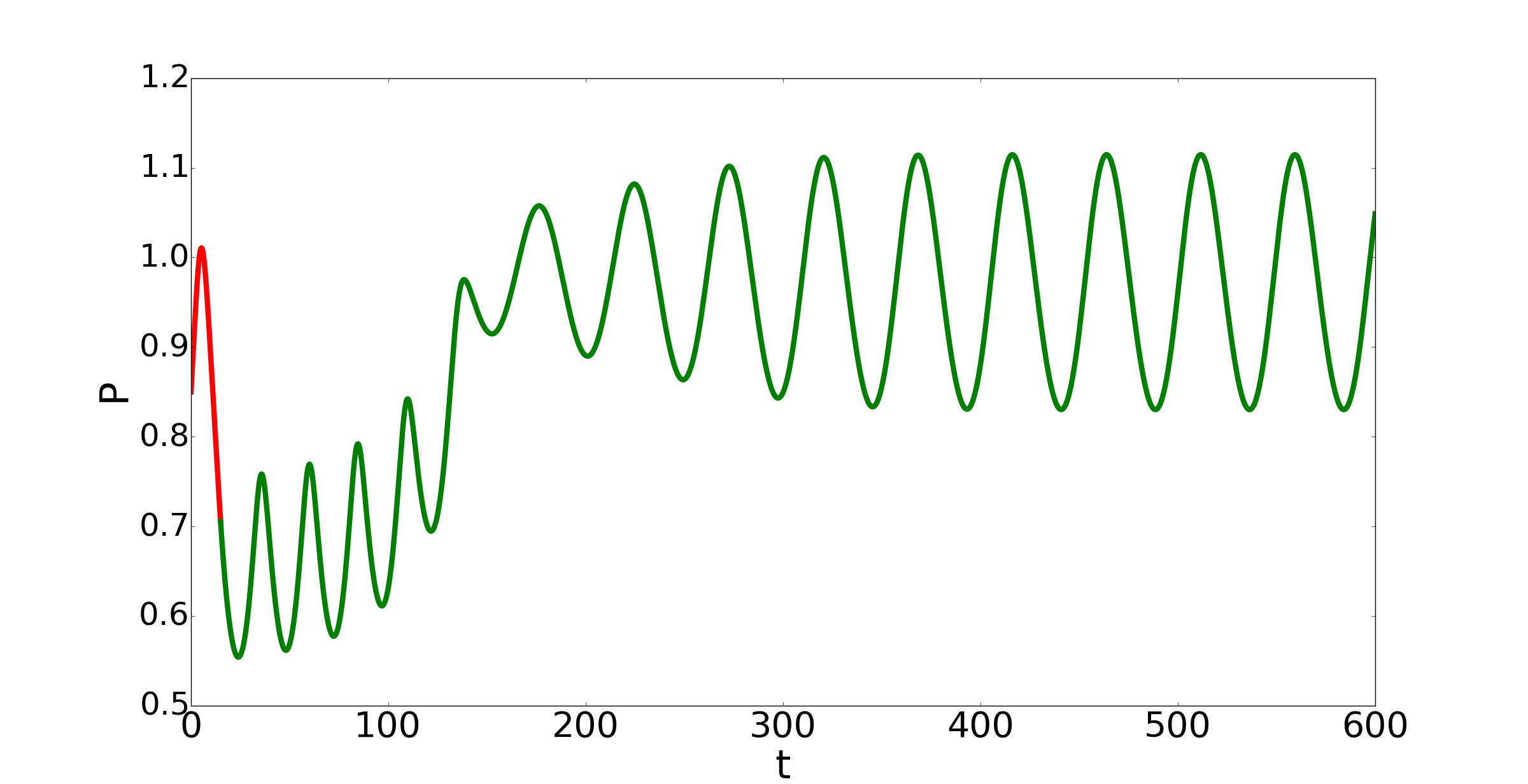} %
		}\qquad		
	\end{subfigure}\hspace*{5cm}
	\begin{subfigure}[h!]{0.2\textwidth}
		\centering
		\subcaptionbox{} 
		{
			\includegraphics[width=2.5\textwidth,height=5cm]{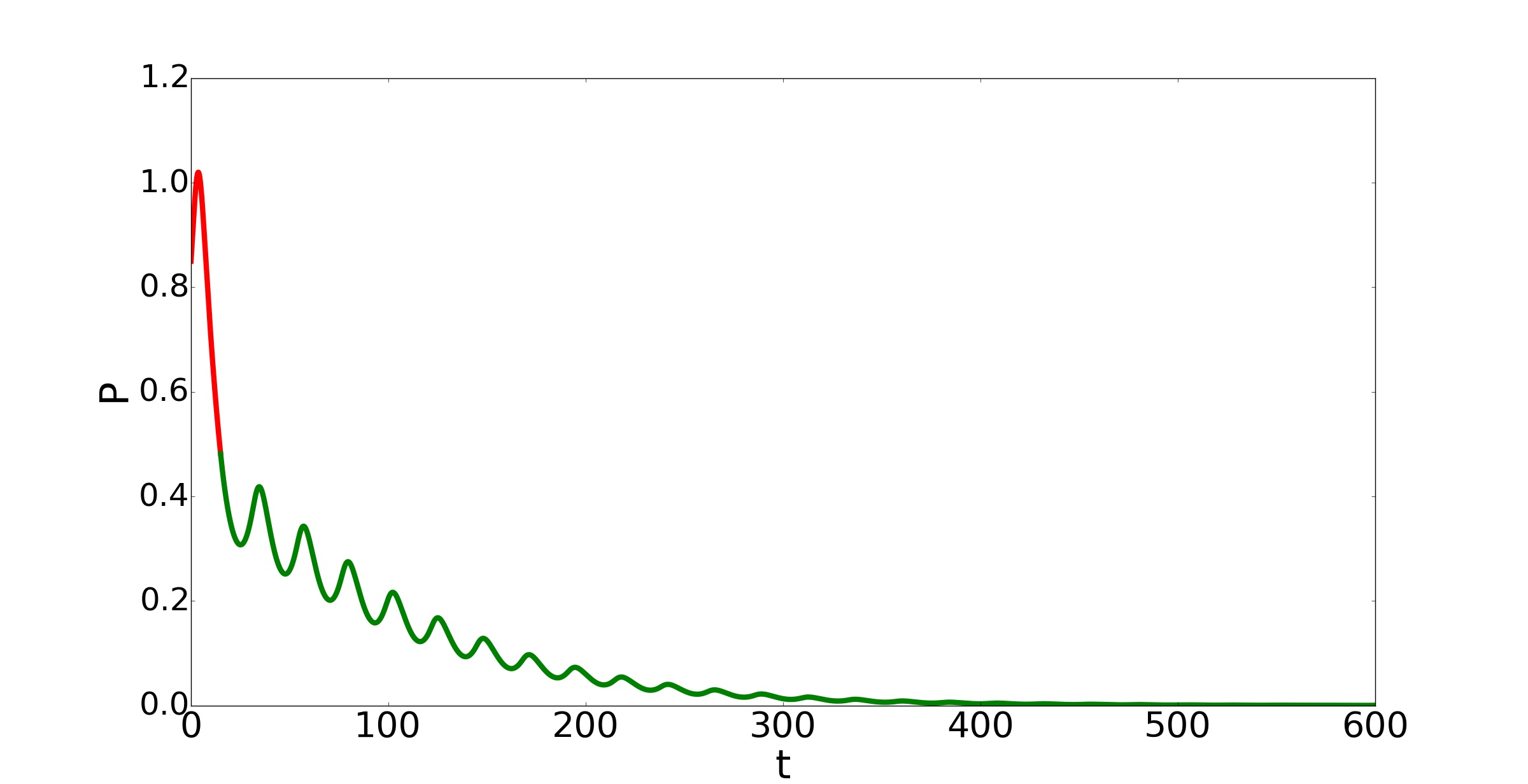}%
		}\qquad	
	\end{subfigure} 
	\caption{{\bf Effect of consumer induction:} (a,b) Red and green curves indicate, respectively, the predator behavior before and after the system is perturbed. The perturbation is done with an addition of the consumer population density of magnitude 0.01 in the consumer growth equation of the model. (0.5,0.61,0.85) and (0.5,0.91,0.85) are the initial points of (a,b) respectively. Parameter values correspond to Table \ref{T1}.}
	\label{pert}		
\end{figure*}

\subsection{A possible restoration mechanism}
\par $\;$ $\;$ It is earlier shown that predator extinction may be avoided in the food chain by supplying additional predators to the system from outside \cite{dhamala1999controlling}. Species extinction can also be averted in this system by supplying additional consumers to maintain predator's growth (see Fig. \ref{pert}). The required number of consumers/predators for the survival of predator species varies with the initial population densities, that is, the location of the basin. Fig.~\ref{pert} shows that a small number of additional consumers can prevent predator extinction where the initial population is considered from the location with multiple modes. However, the same number of consumers cannot protect the predator from its death when the initial population is considered from a unimodal site of the basin, which needs much higher additional consumers. This trend is maintained throughout the basin. 

\bibliographystyle{unsrt}
\bibliography{Ref}
\end{document}